\documentclass[a4, 10pt, usenatbib]{mn2e}

\usepackage[dvips]{graphicx}

\usepackage{amsmath, amssymb}

\usepackage{times}

\usepackage{natbib}
\usepackage{aas_macros}

\def\Vec#1{\mbox{\boldmath $#1$}}

\def\P#1#2{\dfrac{\partial #1}{\partial #2}}
\def\PP#1#2{\dfrac{\partial^2 #1}{\partial #2^2}}

\title[Counter effects of flows and magnetic fields]
{Counter effects of meridional flows and
magnetic fields in stationary axisymmetric
self-gravitating barotropes under the ideal MHD 
approximation: Clear examples --- toroidal configurations}

\author[K. Fujisawa et al.]{Kotaro
Fujisawa\thanks{E-mail: fujisawa@ea.c.u-tokyo.ac.jp} $^\mathrm{1}$,
Rohta Takahashi$^\mathrm{2,3}$, 
Shijun Yoshida$^\mathrm{4}$
and 
Yoshiharu Eriguchi$^\mathrm{1}$ \\
$^\mathrm{1}$ Department of Earth Science and Astronomy,
Graduate School of Arts and Sciences, University of Tokyo,
Komaba, Meguro-ku, Tokyo 153-8902, Japan \\
$^\mathrm{2}$ Department of Natural and Physical Sciences, 
Tomakomai National College of Technology, Tomakomai 059-1275, Japan \\
$^\mathrm{3}$ High Energy Astrophysics Laboratory, 
The Institute of Physical and Chemical Research (RIKEN), Saitama 351-0198, Japan \\
$^\mathrm{4}$ Astronomical Institute, Tohoku University, Sendai 980-8578, Japan
}

\begin{document}

\date{Accepted 2013 February 8, Received 2013 January 25; in original form 2012 December 4}

\pagerange{\pageref{firstpage}--\pageref{lastpage}} \pubyear{2013}

\maketitle

\label{firstpage}

\begin{abstract}

We obtain the general forms for the current density
and the vorticity from the integrability conditions of the basic 
equations which govern {\it the stationary states of axisymmetric
magnetized self-gravitating barotropic objects with meridional 
flows under the ideal MHD approximation}. As seen from the 
stationary condition equations  
for such bodies, the presence of the 
meridional flows and that of the poloidal magnetic fields act 
oppositely on the internal structures. 

The different actions of these two physical quantities,  
the meridional flows and the poloidal magnetic fields, could
be clearly seen through stationary structures of 
the {\it toroidal} gaseous configurations around 
central point masses in the framework of Newtonian gravity 
because the effects of the two physical quantities can be seen 
in an amplified way for toroidal systems compared to
those for spheroidal stars. 
The meridional flows make the structures more compact,
i.e. the widths of toroids thinner,  while the poloidal magnetic 
fields are apt to elongate the density contours
in a certain direction depending on the situation. 
Therefore the simultaneous presence of the internal 
flows and the magnetic fields would work as if 
there were no such different actions within and
around the stationary gaseous objects such as 
axisymmetric magnetized toroids with internal motions 
around central compact objects under the ideal MHD 
approximation, although these two quantities might 
exist in real systems. 

\end{abstract}

\begin{keywords}
 stars: magnetic fields -- rotation -- 
Physical Data and Processes:
 black hole physics -- accretion, accretion discs
\end{keywords}

 \section{Introductory analysis and motivation}
\label{Sec:intro}

\subsection{Theoretical treatment of stationary states of 
axisymmetric magnetized self-gravitating barotropes
under the ideal MHD approximation}
\label{Sec:1.1}

In this paper, we give new expressions for the current density and 
the vorticity vector inside the    
stationary and axisymmetric magnetized self-gravitating barotropes 
with internal gaseous motions under the ideal magnetohydrodynamics (MHD) approximation. 
Although stationary states of the magnetized self-gravitating barotropes without 
internal flows have been investigated in many old papers 
 (e.g., \citealt{Chandrasekhar_Fermi_1953}; \citealt{Lust_Schluter_1954}; 
\citealt{Ferraro_1954}; \citealt{Gjellestad_1954}; 
\citealt{Roberts_1955}; \citealt{Chandrasekhar_1956a}; 
\citealt{Chandrasekhar_1956b}; \citealt{Chandrasekhar_Prendergast_1956}; 
\citealt{Prendergast_1956}; \citealt{Sykes_1957}; 
\citealt{Woltjer_1959a}; \citealt{Woltjer_1959b}; 
\citealt{Woltjer_1960}; \citealt{Ostriker_Hartwick_1968})
and in recent papers
(e.g., \citealt{Tomimura_Eriguchi_2005}; 
\citealt{Yoshida_Eriguchi_2006}; \citealt{Yoshida_Yoshida_Eriguchi_2006}; 
\citealt{Otani_Takahashi_Eriguchi_2009}; 
\citealt{Fujisawa_Yoshida_Eriguchi_2012}), the effects of the internal flows 
on structures of magnetized self-gravitating barotropes have barely studied so far. 

Dynamic equilibrium equations for {\it stationary states of  
magnetized self-gravitating bodies with internal flows} 
are given by 
\begin{eqnarray}
   {1 \over \rho} \nabla p = - \nabla \phi_g 
     - \nabla \phi_c 
     -{1 \over 2} \nabla|\Vec{v}|^2
     + \Vec{v} \times \Vec{\omega} + {1 \over c \rho} \Vec{j} 
       \times \Vec{B} \ , 
\label{Eq:eular_eq}
\end{eqnarray}
where $\rho$,  $p$, $\phi_g$, $\phi_c$, $\Vec{v}$, $\Vec{\omega}$, $c$, $\Vec{j}$, and 
$\Vec{B}$ are 
the density,  the pressure, the gravitational potential of the gaseous body, 
the gravitational potential of external objects, the fluid velocity, the vorticity vector, 
the speed of light, the current density, and the magnetic field, 
respectively. In this study, the gravitational potential of external objects is 
assumed to be given by 
\begin{eqnarray}
 \phi_c  = -  {G M_c \over r} \, ,
\end{eqnarray}
where $M_c$, $G$, and $r$ denote the mass of the central external 
object, the gravitational constant, and the distance from the central object, 
respectively.
Using the expressions for the current density and the vorticity
vector inside the stationary, axisymmetric and infinitely conducting barotropes  
(for  details, see Appendix \ref{app:integrability}), 
we may simplify the forth term in the right-hand side of Eq. (\ref{Eq:eular_eq}) as   
{\scriptsize
\begin{eqnarray}
  \Vec{v} \times \Vec{\omega} =
     \left\{
        \begin{array} {ll}  
          \Omega(\Psi) \nabla (R v_{\varphi}) 
              + \displaystyle{1 \over 4 \pi c \rho} 
                \displaystyle{dQ(\Psi) \over d\Psi}
               \Vec{B} \times \Vec{\omega} \ , & 
             (\mbox{for $\Vec{B} \ne 0, \Vec{v}_p \ne 0$}) \\
         \nu(Q) \nabla Q \ , & 
            (\mbox{for  $\Vec{B} = 0, \Vec{v}_p \ne 0$}) \\ 
         {1 \over 2} \nabla (R^2 \Omega^2(R)) 
              + R \Omega^2(R)  \Vec{e}_R  \ , & 
           (\mbox{for $\Vec{B} = 0, \Vec{v}_p = 0$})
       \end{array}
     \right.
\label{Eq:vxw}
\end{eqnarray}
}
where $\Psi$ and $Q$ are the flux and stream functions, respectively 
(see Eqs. (\ref{Eq:stream_func}) and (\ref{Eq:flux})), 
$\Omega$ and $\nu$ are arbitrary functions of the given argument, 
and $R$ and $\Vec{e}_R$ are the radial coordinate and base vector 
for the cylindrical coordinate with respect to the symmetry axis, respectively. 
We describe the details of these arbitrary functions in 
Appendix \ref{app:integrability} and 
Appendix \ref{app:integrability2}.
Here, $\Vec{v}_p$ and $v_\varphi$ stand for the poloidal velocity 
and the toroidal velocity, given by  
\begin{eqnarray}
\Vec{v}_p = {1\over 4\pi\rho}\nabla Q \times \nabla \varphi \,,
\end{eqnarray}
\begin{eqnarray}
     v_{\varphi} = R \Omega 
         + \displaystyle{1 \over 4 \pi c \rho} 
            \displaystyle{dQ(\Psi) \over d \Psi}
                     B_{\varphi} \,,
\end{eqnarray}
with $\varphi$ and $B_{\varphi}$ being the azimuthal angle and the toroidal magnetic field, respectively. 
The stream function $Q$ 
is a function of the flux function $\Psi$ if and only if $\Vec{B} \ne 0$ and $\Vec{v}_p \ne 0$. 
Thus, equi-stream function surfaces  are 
similar to  equi-magnetic flux function surfaces if $\Vec{B}\ne0$ and $\Vec{v}_p\ne0$.
Similarly,  the Lorentz force term appearing in the fifth term in the right-hand side 
of  Eq. (\ref{Eq:eular_eq}) may be rewritten as 
{\scriptsize
\begin{eqnarray}
  {1 \over c \rho} \Vec{j} \times \Vec{B} =
    \left\{ 
      \begin{array}{ll}  
        \nabla (\Omega(\Psi)) R v_{\varphi}
           + \mu(\Psi) \nabla \Psi 
           + \displaystyle{1 \over 4 \pi c \rho} 
        \displaystyle{dQ(\Psi) \over d\Psi}  \Vec{\omega} 
             \times \Vec{B} \ ,  &
          (\mbox{for $\Vec{B} \ne 0$}) \\
        0  \ , & 
          (\mbox{for $\Vec{B} = 0$}) \\
     \end{array}
   \right.
\label{Eq:lorentz_force}
\end{eqnarray}
}
where $\mu(\Psi)$ is an arbitrary function.
From Eqs. (\ref{Eq:eular_eq})--(\ref{Eq:lorentz_force}),  
we obtain Bernoulli's equations for the present situation:  
{\scriptsize
\begin{eqnarray}
  \int {dp \over \rho} = 
    \left\{
      \begin{array}{l}
      - \phi_g  +  \displaystyle{G M_c \over r} 
       -{1 \over 2} (|\Vec{v}_p|^2  + v_{\varphi}^2)
         +  R v_{\varphi} \Omega(\Psi)  
         + \int^{\Psi} \mu(\Psi) d \Psi + C \ ,  
         \label{Eq:first_int}
                       \\
       \qquad      (\mbox{for $\Vec{B} \ne 0$}) \\  
       - \phi_g + \displaystyle{G M_c \over r}
          -{1 \over 2} (|\Vec{v}_p|^2 + v_{\varphi}^2)
          + \int^{Q} \nu(Q) d Q + C \ ,  \\
       \qquad     (\mbox{for $\Vec{B} = 0, 
           \Vec{v}_p \ne 0$}) \\
        - \phi_g + \displaystyle{G M_c \over r}
           + \int^R R \Omega^2 dR + C \ , \\
       \qquad       (\mbox{for $\Vec{B} = 0,   \Vec{v}_p= 0$}) \\
    \end{array}
  \right.
\label{eq:first_integral}
\end{eqnarray}
}
where $C$ is an integration constant. 
It should be emphasized that these three expressions cannot be united  to a single  
expression as can be understood from Eqs. (\ref{Eq:vxw}) and (\ref{Eq:lorentz_force}). 
Note that in this study, we do not 
consider the case of $\Vec{B}_p=0$ and $B_\varphi\ne0$, for which Bernoulli's equations 
differ from Eq. (\ref{eq:first_integral}). 

By comparing Bernoulli's equation for the case of  $\Vec{B} = 0$ and $\Vec{v}_p = 0$ with  
that for the case of $\Vec{B} = 0$ and $\Vec{v}_p \ne 0$, we may expect that  
{\it the presence of the meridional flows} tends to decrease the volume occupied by the fluid. 
This is because the kinetic energy of the meridional flow contributes to 
the fluid as 
{\it positive ram pressures} (dynamic pressure), which result in reducing 
the fluid pressure and consequently decreasing  the 
density of stationary fluid objects.
In contrast to the effect of the meridional flow, for magnetized equilibrium 
configurations with $\Vec{v}_p = 0$ whose magnetic fields are generated by 
positive toroidal currents, the poloidal magnetic field is apt to 
expand the fluid  
region to the direction perpendicular  to the symmetry axis like 
the centrifugal force (see, e.g.  old papers 
\citealt{Ferraro_1954}; \citealt{Gjellestad_1954};
\citealt{Roberts_1955}; and recent papers 
\citealt{Tomimura_Eriguchi_2005};
\citealt{Yoshida_Eriguchi_2006};
\citealt{Yoshida_Yoshida_Eriguchi_2006};
\citealt{Fujisawa_Yoshida_Eriguchi_2012}). 
Therefore, simultaneous presence of the meridional flows and 
the poloidal magnetic fields could result in almost no changes 
in the matter distributions. So far, such effects  
have not been investigated because stationary states of 
axisymmetric magnetized barotropes with meridional 
flows have not been obtained. Thus it is one of the purposes in 
this study to show the above statements could be proved 
to hold in some numerical examples.

\subsection{Toroids: Best astrophysical systems in which 
both meridional flows and magnetic fields would work 
simultaneously but differently}

In order to show clearly the oppositely working effects 
of the above mentioned two quantities, meridional flows and poloidal magnetic fields, 
we will show numerical results of stationary
configurations of axisymmetric magnetized self-gravitating 
{\it toroidal} barotropes with meridional flows which 
locate around central point masses in the framework of Newtonian 
gravity under the ideal MHD approximation. 
For toroidal configurations, matter distribution
changes in wider space would be expected compared to size 
changes of spheroidal objects because toroids could change 
their shapes to {\it two} opposite directions, i.e.  to the outside 
direction  and to the inside direction of the toroids. 
Thus we will solve stationary states of axisymmetric magnetized 
barotropic {\it toroids}  with meridional flows under the ideal
MHD approximation 
and clarify the oppositely working effects explicitly
in this study.

Concerning self-gravitating toroidal configurations or disks, 
we need to take recent results of fully general relativistic 
(GR) numerical simulations into account. 
These simulations show 
that a few-solar-mass black hole 
and a highly dense toroid whose maximum 
density can reach 
$10^{10}$ - $10^{11} \mathrm{g/cm}^3$ around the black hole 
could be formed after merging of binary neutron 
stars (\citealt{Shibata_Uryu_2000};
\citealt{Shibata_Taniguchi_Uryu_2003};
\citealt{Shibata_Taniguchi_Uryu_2005};
\citealt{Kiuchi_Sekiguchi_Shibata_Taniguchi_2009};
\citealt{Hotokezaka_et_al_2011}),
after merging  of a neutron star and a black hole in binary systems 
(\citealt{Shibata_Uryu_2006};\citealt{Shibata_Uryu_2007};
\citealt{Shibata_Taniguchi_2008};
\citealt{Kyutoku_Shibata_Taniguchi_2010};
\citealt{Kyutoku_et_al_2011})
or after collapsing of a supermassive rotating star
(\citealt{Shibata_2000};
\citealt{Shibata_Shapiro_2002};
\citealt{Shibata_2003};
\citealt{Sekiguchi_Shibata_2004};
\citealt{Sekiguchi_Shibata_2007};
\citealt{Sekiguchi_Shibata_2011}).
Therefore, dense toroids and central compact objects could be 
formed after collapsing or merging of compact objects.
Similar kinds of systems with {\it magnetic  fields} have also
been investigated by several groups
(e.g. \citealt{Narayan_et_al_2001}; \citealt{Shibata_Sekiguchi_2005};
\citealt{Duez_et_al_2006}; 
\citealt{Shibata_Sekiguchi_Takahashi_2007}).
Although, in order to understand the origin and dynamical 
formation processes of these systems, we must take into 
account  many realistic physics  and compute stationary
configurations with magnetic fields in GR, nobody has yet succeeded
in solving stationary states both with 
poloidal and toroidal
magnetic fields {\it in GR at present}. 
Therefore, we explore such stationary states of axisymmetric
magnetized barotropic systems in the framework of Newtonian 
gravity. Although there were many papers to obtain 
magnetized stationary states of disks/toroids only with 
poloidal magnetic fields
(e.g. \citealt{Bisnovatyi-Kogan_Blinnikov_1972}; 
\citealt{Bisnovatyi-Kogan_Seidov_1985};
\citealt{Baureis_et_al_1989}; \citealt{Li_Shu_1996}) 
and disks/toroids  only with toroidal magnetic fields 
inside the disks/toroids
(e.g. \citealt{Okada_Fukue_Matsumoto_1989}; 
\citealt{Banerjee_et_al_1995};
\citealt{Ghanbari_Abbassi_2004}),
no solutions both with poloidal and toroidal components of 
magnetic fields have been obtained yet.
This is because it has been difficult  to solve the 
Grad-Shafranov equation as well as the equations of motion 
consistently by some
means. Concerning this type of systems, the most general 
formulation was derived by \cite{Lovelace_et_al_1986} 
systematically.
However, \cite{Lovelace_et_al_1986} computed solutions only 
with poloidal magnetic fields for non-self-gravitating disks.

Recently, \cite{Otani_Takahashi_Eriguchi_2009} have obtained
magnetized self-gravitating equilibrium states both with 
poloidal and toroidal magnetic fields self-consistently 
in the framework of Newtonian gravity. Their method is based 
on \cite{Tomimura_Eriguchi_2005}. In this study we have 
extended the method employed in 
\cite{Otani_Takahashi_Eriguchi_2009} to the most
general configurations for the stationary states  of
axisymmetric magnetized barotropic toroids with
meridional flows under the ideal MHD approximation  
and obtained sequences
of stationary states. Comparing these results
with those of non-magnetized toroids without meridional flows
(e.g. \citealt{Ostriker_1964};\citealt{Wong_1974}), 
with those of magnetized toroids without meridional flows
(e.g. \citealt{Otani_Takahashi_Eriguchi_2009})
or with those of non-magnetized toroids with meridional flows
(e.g. \citealt{Eriguchi_Mueller_Hachisu_1986}), 
we will be able to clearly see the effect of the
presence of both physical
quantities, i.e. meridional flows and magnetic
fields as explained in the previous subsection. 

\section{Brief description of the problem}

In this study, as mentioned, we investigate {\it stationary configurations
of axisymmetric magnetized polytropic toroids with internal fluid 
motions}. We consider {\it inviscid and 
infinitely conductive
toroids with equatorial symmetry} located around
{\it central point masses} in the framework of
{\it Newtonian gravity}. Since a similar problem, but without 
meridional flows, was already treated by \cite{Otani_Takahashi_Eriguchi_2009} 
and our strategy is basically the same as theirs, we briefly summarize 
 the basic equations, boundary conditions, and solving scheme. 
We describe the details of physical quantities 
and dimensionless forms in Appendix 
\ref{app:physical_quantities} 
and the numerical scheme in Appendix \ref{app:scheme_method}.

Continuity and pressure balance equations for stationary states are respectively 
given by  
\begin{eqnarray}
  \nabla \cdot (\rho \Vec{v}) = 0 \,,
    \label{Eq:continuity}
\end{eqnarray}
{\small
\begin{eqnarray}
  \frac{1}{\rho} \nabla p = - \nabla \phi_g + 
     \nabla  \left(\frac{G M_c}{r}\right)
      - \frac{1}{2} \nabla |\Vec{v}^2|
      + \Vec{v} \times \Vec{\omega} + 
      \frac{1}{\rho}\left(\frac{\Vec{j}}{c} 
         \times \Vec{B}  \right) \,.
         \label{Eq:motion_tor}
\end{eqnarray}
}
The Poisson equation for $\phi_g$ is 
\begin{eqnarray}
  \Delta \phi_g = 4 \pi G \rho \, . 
      \label{Eq:Poisson_gpote}
\end{eqnarray}
We make use of the polytropic equation of state  
\begin{eqnarray}
    p = K_0 \rho^{1 + \frac{1}{N}} \, ,
\end{eqnarray}
where $K_0$ and $N$ are a constant and
the polytropic index, respectively.

Maxwell equations for stationary electromagnetic fields are 
\begin{eqnarray}
  \nabla \cdot \Vec{B} & = & 0 \, , 
    \label{Eq:div_B} \\
  \nabla \times \Vec{E} & = & 0 \, , 
    \label{Eq:rot_E} \\
  \nabla \times \Vec{B} & = & 4\pi \frac{\Vec{j}}{c} \, ,
    \label{Eq:rot_H} 
\end{eqnarray}
where $\Vec{E}$ is the electric field, which is determined by 
the perfect conductivity condition 
\begin{eqnarray}
 \Vec{E} + \frac{\Vec{v}}{c} \times \Vec{B} =0  \, . 
    \label{Eq:ideal_MHD} 
\end{eqnarray}
The magnetic flux function $\Psi$ is, in terms of the azimuthal component 
of the vector potential $A_\varphi$, defined by 
\begin{eqnarray}
  \Psi = R A_{\varphi} \ .
\label{Eq:Psi_Aphi}
\end{eqnarray}
From the azimuthal component  of the Maxwell equation (\ref{Eq:rot_H}), 
we obtain 
\begin{eqnarray}
  R \P{\Psi}{R} \left(\frac{1}{R} \P{\Psi}{R} \right) 
    + \PP{\Psi}{z} = - 4 \pi R \frac{j_\varphi}{c} \ ,
\label{Eq:GS}
\end{eqnarray}
where the cylindrical coordinates $(R,\varphi, z)$ have been used. 
This equation is equivalent to the so called Grad-Shafranov equation 
for this problem,  but we treat the right-hand side of Eq. (\ref{Eq:GS}) as the source 
term of the differential operator in the left-hand side of Eq. (\ref{Eq:GS}) even though 
$j_\varphi$ includes the term proportional to  the left-hand side of Eq. (\ref{Eq:GS}) 
if $\Vec{v}_p\ne0$. Explicit expression for $j_\varphi$ is given in Eq. (\ref{Eq:current}). 
In order to solve Eq. (\ref{Eq:GS}) numerically, it is useful to transform Eq. (\ref{Eq:GS}) 
into 
\begin{eqnarray}
 \Delta (A_\varphi \sin \varphi) 
    = - 4\pi \frac{j_\varphi}{c} \sin \varphi \,,
  \label{Eq:2_GS}
\end{eqnarray}
where $\Delta$ denotes the Laplacian 
(see, e.g.,  \citealt{Tomimura_Eriguchi_2005}
and \citealt{Otani_Takahashi_Eriguchi_2009}). 

Because isolated mass and current densities are considered in this study, 
boundary conditions at infinity for the gravitational potential $\phi_g$ 
and the vector potential $A_{\varphi}$ are respectively given by 
\begin{eqnarray}
   \phi_g = {\cal O} \left(\frac{1}{r}\right) \, , 
        \hspace{10pt}
  A_{\varphi} = {\cal O} \left(\frac{1}{r}\right) \, . 
  \label{Eq:BC}
\end{eqnarray}
Solutions for the present problem are obtained 
by solving the pressure balance equation and  the two Poisson equations
for $\phi_g$ and $A_{\varphi}\sin \varphi$ with the boundary 
condition (\ref{Eq:BC}). In order to impose the boundary condition (\ref{Eq:BC}) 
automatically, in this study, the two Poisson equations (\ref{Eq:Poisson_gpote}) and (\ref{Eq:2_GS}) are 
converted into the two integral equations, given by 
\begin{eqnarray}
 \phi_g & = & - G \int \frac{\rho(\Vec{r}')}{|\Vec{r}-\Vec{r}'|} \, 
  d^3\Vec{r}' \ , 
   \label{Eq:2_Poisson_int} \\
 A_\varphi(r, \theta) \sin \varphi & = & \frac{1}{c}\int 
   \frac{j_\varphi(\Vec{r}')}{|\Vec{r}-\Vec{r}'|} \sin \varphi' 
   \, d^3 \Vec{r}' \ . 
  \label{Eq:A_phi}
\end{eqnarray}
Integrating Eq. (\ref{Eq:motion_tor}), we obtain Eq. (\ref{eq:first_integral}). 
Eqs. (\ref{eq:first_integral}), (\ref{Eq:2_Poisson_int}), 
and (\ref{Eq:A_phi}) are the master equations for the present problem, which 
are solved with a variant of the numerical scheme used by 
\cite{Otani_Takahashi_Eriguchi_2009} 
(for details, see Appendix \ref{app:scheme_method}).

In general, the toroid cannot approach indefinitely to the central
object because gravitational effects of the central object can unboundedly 
increase as the distance from the central object to the toroid decreases 
and any forces counteracting the gravity cannot stanch the matter flow 
shedding from the inner edge of the toroid if their distance is shorter than 
some critical value. In the present numerical scheme to obtain magnetized 
toroids, the distance from the central object to the inner edge of the toroid (or the width of the toroid) is characterized by 
a dimensionless parameter $q$, defined by 
\begin{eqnarray}
 q \equiv \frac{R_{\rm{inner}}}{R_{\rm{outer}}},
\label{Eq:q}
\end{eqnarray}
where $R_{\rm{inner}}$ and $R_{\rm{outer}}$ are the shortest and 
the longest distances from the symmetry axis to the toroid, respectively.  
In terms of $q$, this disappearance property of the equilibrium sates 
describes as the existence of $q_c$ such that there is 
no stationary solution of the toroid for $q  < q_{\rm c}$. Note that the value of 
$q_c$ depends on what parameters characterizing equilibrium sequences 
keep constant when the value of $q$ changes. 
Following \cite{Otani_Takahashi_Eriguchi_2009}, we call 
equilibrium solutions characterized by $q=q_c$ the critical 
configuration or the critical state. The distance from the symmetry axis  to the inner edge of 
the toroid for the critical configuration is named 
the {\it critical distance}. In this study, we focus only on the critical 
configurations.

\cite{Otani_Takahashi_Eriguchi_2009}  investigated the critical 
configuration for the magnetized toroids without meridional flows 
and found the following properties.  (i) The critical configuration 
features cusp-like structures at the inner edge of the toroids. 
(ii) The critical configuration rotates very slowly. 
This implies that the critical toroids are
mainly sustained by the balance among the magnetic 
forces, the gravity of the central objects and
the pressure gradients. 
(iii) The critical distances are almost independent of the 
mass ratio of the toroids to the central objects. (iv) 
The critical distances are much larger than 
$6 G M_c/c^2$, the radius of the innermost stable circular 
orbit for the Schwarzschild black hole with the gravitational mass $M_c$.  
This means that making use of Newtonian gravity is reasonable to 
investigate structures of the toroids considered 
in \cite{Otani_Takahashi_Eriguchi_2009}.

\section{Numerical results}
\label{Sec:numerical_results}

Following \cite{Otani_Takahashi_Eriguchi_2009}, 
we consider two polytropic  indices $N=1.5$ and $N=3$ only in the present study. 
As for the arbitrary functions, which need to be specified to obtain particular solutions, 
we employ the same functional forms as those used 
in \cite{Yoshida_Yoshida_Eriguchi_2006} and
\cite{Otani_Takahashi_Eriguchi_2009} except for the toroidal current function 
$\mu$. For the toroidal current function $\mu$, we choose the same functional form as that used 
in \cite{Fujisawa_Yoshida_Eriguchi_2012} in order for the inner edge of the toroid to have stronger 
magnetic fields.  
As for the stream function $Q'$, which does not appear in \cite{Otani_Takahashi_Eriguchi_2009}, 
a similar functional form to that of the poloidal current function $\kappa$ is employed. 
Details of the functional forms of the arbitrary functions are collected in 
Appendix \ref{app:arbitrary_functions}.

As mentioned before, recent numerical simulations performed with numerical relativity show that 
geometrically thick toroids rotating around black holes form after 
mergers of neutron star binaries, mergers of black hole-neutron star binaries, 
or collapses of  supermassive rotating stars
(e.g. \citealt{Sekiguchi_Shibata_2007}; 
\citealt{Sekiguchi_Shibata_2011}). Typical values of  physical quantities of such 
black hole-toroid systems are given by $M_t = 1.0 \times 10^{-1} M_\odot$, 
$M_c = 5.0  M_\odot$  ($M_t / M_c = 2.0 \times 10^{-2}$), and 
$\rho_{\rm{c}} = 1.0 \times 10^{11} \rm{g/cm^3}$, where $M_t $ and 
$\rho_{\rm{c}}$ are the mass and the maximum density of the toroid, 
respectively. Since these black hole-toroid systems are of large significance 
in high energy astrophysics, we focus on  the models characterized 
by the mass ratio $M_t / M_c = 2.0 \times 10^{-2}$ and use these values of 
the mass of the central object and the maximum density of the toroid to estimate values of 
other physical quantities with physical dimension, e.g. $|\Vec{B}|$ and $R_{inner}$.  

To check numerical accuracies of the stationary configurations obtained 
in this study,  we estimate values of a virial relation, which vanishes for
exact stationary solutions: 
\begin{eqnarray}
  {\rm VC} =  \left|\frac{2 T + W + 
        3 \Pi + \mathfrak{M}}{W}\right| \ ,
\end{eqnarray}
where $T, W, \Pi$, and $\mathfrak{M}$ are the
kinetic energy, the gravitational energy, the volume
integral of the pressure and the magnetic energy,
respectively
(for details, see Appendix \ref{app:physical_quantities}). 
As shown later, all the stationary configurations obtained in this study 
are within acceptable accuracy (for details,  see Appendix \ref{app:Numerical}).

\subsection{Widening of the widths of toroids: Effect of the 
localized poloidal magnetic fields}

\begin{table*}
 {\small
\begin{tabular}{ccccccccccccc}
\hline
$m$ & $q_c$ & $\hat{\mu}_0$ & $r_e$(cm) &  $|\hat{W}|$ 
& $ \mathfrak{M}/|W|$ & $U/|W|$ &  VC \\
 \hline
&&&&\hspace{-20pt} $N=1.5$ \\
\hline

0.5  & 0.680 & 6.960E+00 & 2.984E+07 & 2.573E-02 & 6.509E-01 &
 1.164E-01 & 2.763E-05 \\

0.3 & 0.650 & 4.700E+00 & 2.791E+07 & 3.893E-02 & 6.421E-01 &
 1.193E-01 & 2.713E-05 \\

0.0 & 0.598 & 2.984E+00 & 2.518E+07 & 7.401E-02 & 6.272E-01 &
 1.243E-01 & 2.610E-05 \\  
 
-0.1 & 0.578 & 2.661E+00 & 2.432E+07 & 9.226E-02 & 6.218E-01 &
 1.261E-01 & 2.575E-05 \\
 
-0.3 & 0.533 & 2.238E+00 & 2.264E+07 & 1.449E-01 & 6.100E-01 &
 1.300E-01 & 2.525E-05 \\
 
-0.5 & 0.481 & 2.033E+00 & 2.105E+07 & 2.312E-01 & 5.966E-01 &
 1.345E-01 & 2.444E-05 \\
 
-0.7 & 0.419 & 1.996E+00 & 1.954E+07 & 3.754E-01 & 5.812E-01 &
 1.396E-01 & 2.381E-05 \\
 
-0.9 & 0.344 & 2.117E+00 & 1.812E+07 & 6.201E-01 & 5.631E-01 &
 1.456E-01 & 2.329E-05 \\
 
-1.1 & 0.252 & 2.408E+00 & 1.683E+07 & 1.036E+00 & 5.415E-01 &
 1.528E-01 & 2.274E-05 \\

-1.4  & 0.067 & 2.956E+00 & 1.564E+07 & 1.923E+00 & 5.111E-01 &
 1.630E-01 & 2.471E-05 \\ 

\hline
&&&&\hspace{-20pt} $N=3$ \\
\hline

 0.5 & 0.720 & 1.159E+01 & 4.653E+07 & 1.747E-03 & 6.910E-01 &
 1.030E-01 & 6.433E-05 \\

 0.3 & 0.687 & 6.031E+00 & 4.243E+07 & 3.080E-03 & 6.807E-01 &
 1.065E-01 &  6.073E-05 \\

 0.0 & 0.628 & 2.742E+00 & 3.656E+07 & 7.732E-03 & 6.639E-01 &
 1.121E-01 &  5.472E-05 \\
  
-0.1 & 0.605 & 2.227E+00 & 3.473E+07 & 1.064E-02 & 6.578E-01 &
 1.141E-01 &  5.279E-05 \\
 
-0.3  & 0.554 & 1.600E+00 & 3.129E+07 & 2.044E-02 & 6.446E-01 &
 1.185E-01 &  4.880E-05 \\
 
-0.5  & 0.495 & 1.289E+00 & 2.814E+07 & 3.993E-02 & 6.297E-01 &
 1.234E-01 &  4.470E-05 \\
 
-0.7  & 0.425 & 1.164E+00 & 2.529E+07 & 7.927E-02 & 6.127E-01 &
 1.291E-01 & 4.064E-05 \\
 
-0.9  & 0.342 & 1.173E+00 & 2.274E+07 & 1.593E-01 & 5.926E-01 &
 1.358E-01 &  3.644E-05 \\
 
-1.1  & 0.242 & 1.298E+00 & 2.057E+07 & 3.195E-01 & 5.678E-01 &
 1.441E-01 &  3.224E-05 \\
 
 -1.4 & 0.057 & 1.467E+00 & 1.901E+07 & 6.864E-01 & 5.261E-01 &
 1.580E-01 &  2.944E-05 \\
 
\hline

\end{tabular}
}
\caption{Physical quantities for the critical configurations 
with $\hat{Q}_0 = 0.0$ (no meridional flow), $\hat{\Omega}_0 = 0.0$ (no rotation)
and $\hat{\kappa}_0 = 4.5$. 
}
\label{Tab:mu}
\end{table*}

\begin{figure*}
\begin{center}
 \includegraphics[scale=1.3]{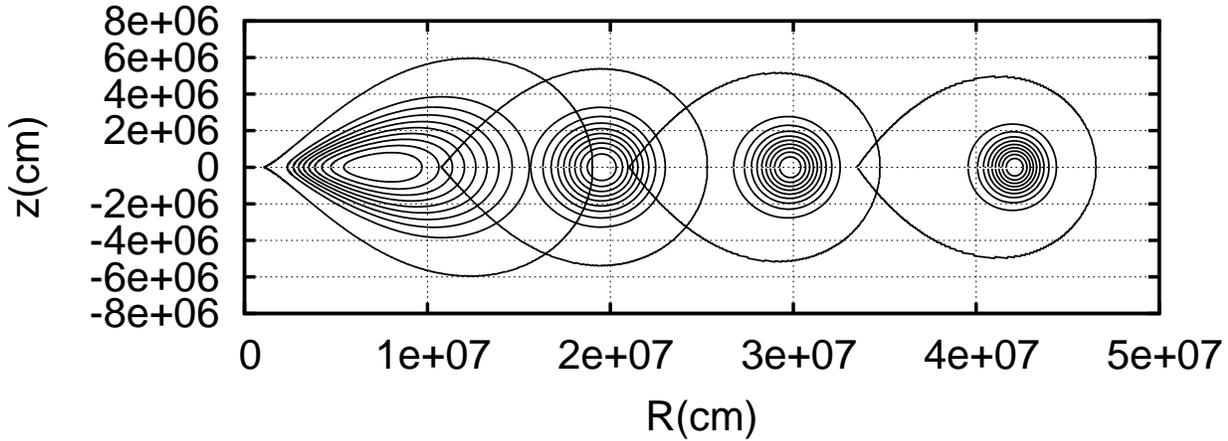}
\end{center}
 \caption{The density contours on the meridional cross section 
for four critical configurations with different values of $m$. From left to right, 
the density contours correspond to the toroids with $m = -1.4$, 
$m=-0.7$, $m=-0.1$, and  $m=0.5$, respectively. Values of 
the other parameters are $ N = 3$, $\hat{Q}_0 = 0$, $\hat{\Omega}_0 = 0$, 
and $\hat{\kappa}_0 = 4.5$. The tear-shaped closure curves with 
a cusp-like structure indicate the surfaces of the toroids. The density difference 
between two adjacent contours is one-tenth of the maximum density. 
It is observed that 
the width of the toroid on the equator becomes wider as values 
of $m$ decrease.
}
\label{fig:tori}
\end{figure*}

\begin{figure*}
  \includegraphics[scale=0.85]{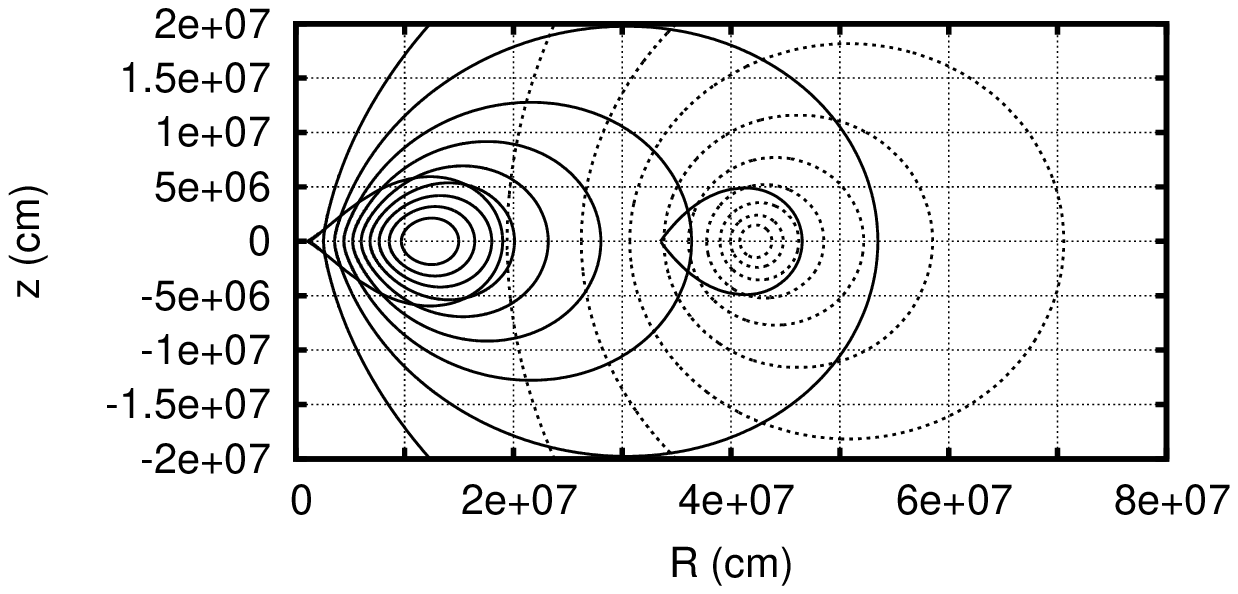} 

\includegraphics[scale=0.75]{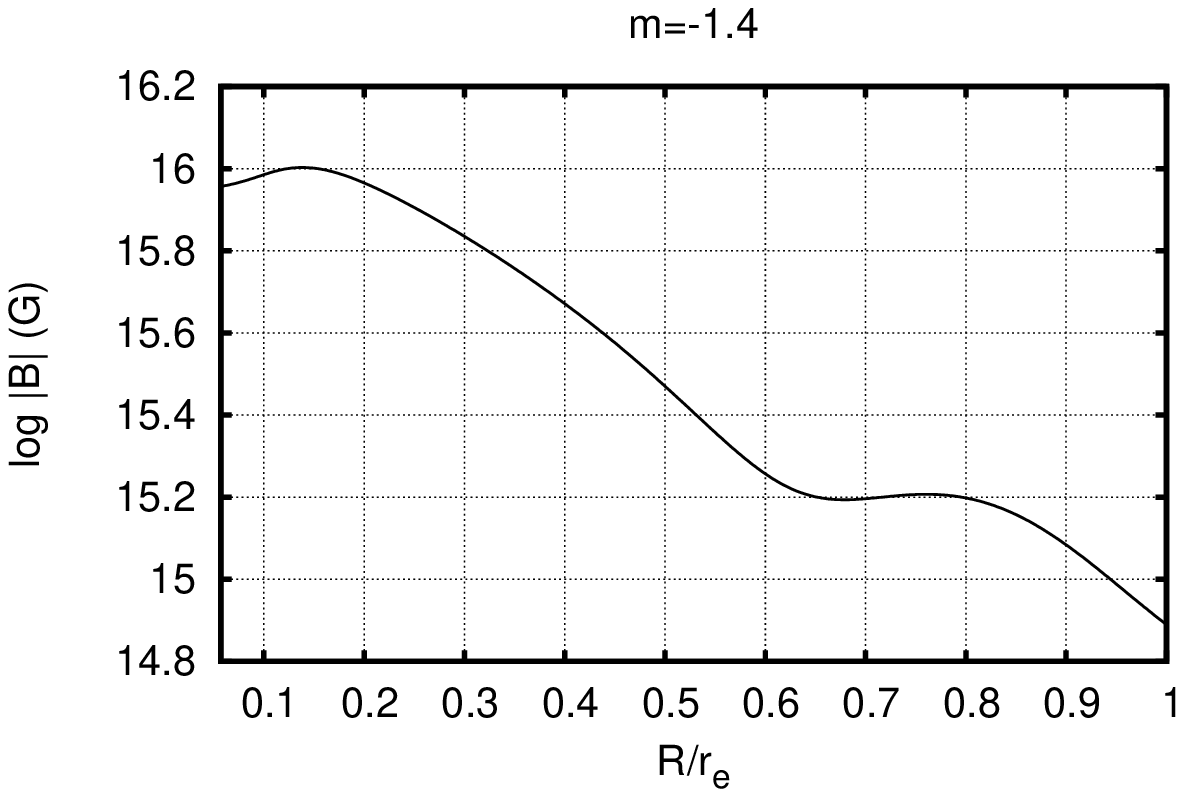}
\includegraphics[scale=0.85]{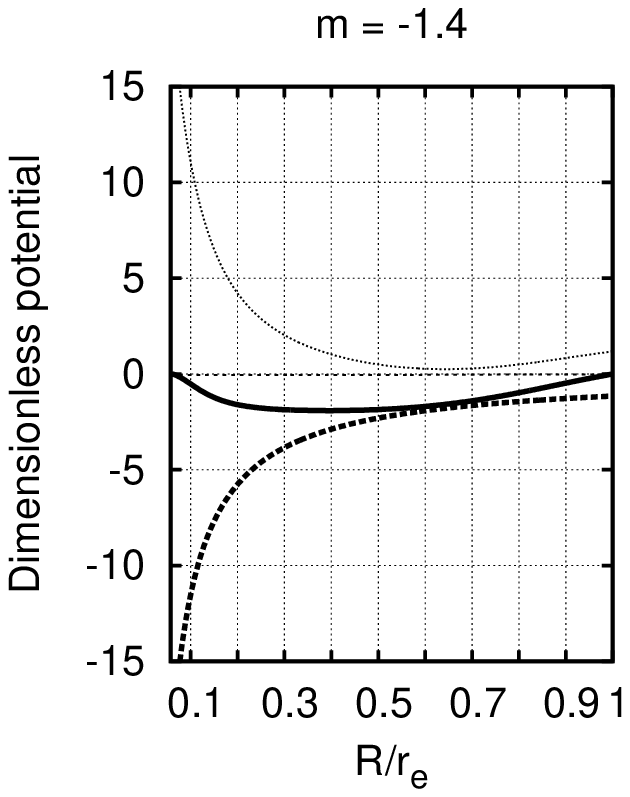}

\includegraphics[scale=0.75]{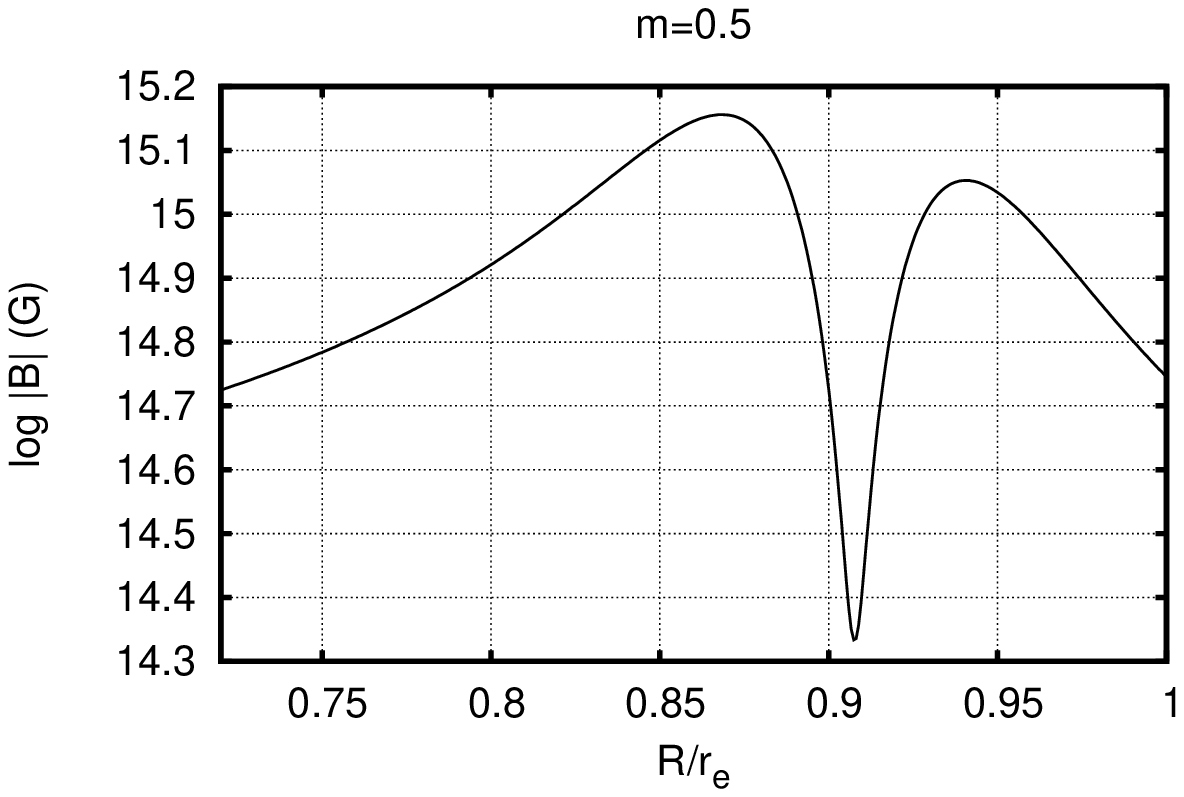}
\includegraphics[scale=0.85]{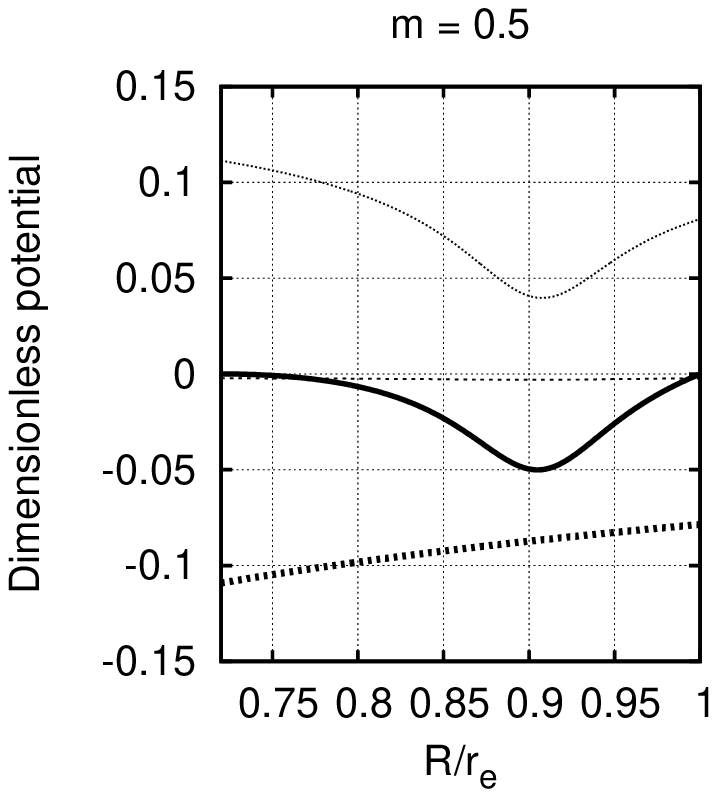}

\caption{Top panel: 
The poloidal magnetic fields on the meridional plane
for the $N=3$ toroids with $m=-1.4$ (solid curves) and $m=0.5$ (dashed curves). 
Values of the other model parameters are the same as those of the models given in 
Fig. \ref{fig:tori}. The tear-shaped closure curves with a cusp-like structure indicate  
the surfaces of the toroids.
Middle left and bottom left panels: 
$\log_{10} |\Vec{B}|$ on the equator for two critical configurations with 
$m = -1.4$ (middle) and $m = 0.5$ (bottom), given as functions of $R/r_e$. 
The horizontal axises range from $R_{inner}/r_e(=q_c)$ to $R_{outer} /r_e(=1)$. 
Middle right and bottom right panels:  
$-\hat{M_c}/ 4\pi\hat{r}$ (thick dashed curve), $\hat{\phi}_g$ 
(thin dashed curve),  $-\int \hat{\mu}(\hat{\Psi})\, d \hat{\hat{\Psi}} - \hat{C}$ 
(dotted curve), and sum of these three potentials (solid curve) on the equator 
for the critical configurations with $m=-1.4$ (middle) and $m=0.5$ (bottom), given 
as functions of $R/r_e$. The horizontal axises range from 
$R_{inner}/r_e(=q_c)$ to $R_{outer} /r_e(=1)$.
}
\label{fig:tori2}
\end{figure*}

The counter effects of the  meridional flows against
the magnetic forces on structures of magnetized toroids would be
clearly seen for toroids with rather widened shapes
due to poloidal magnetic fields. Thus, in this subsection, we will
try to compute magnetized toroids with highly localized 
poloidal magnetic fields because such equilibrium configurations
could be toroids with a rather small value of $q$,
i.e. the width of toroids on the equatorial plane
being rather wide (see e.g. \citealt{Fujisawa_Yoshida_Eriguchi_2012}).

To investigate effects of the localized poloidal magnetic fields on the toroid structures, 
no fluid flow inside the toroid is considered here. Thus, values of the parameters 
$\hat{Q}_0 $, $\hat{\Omega}_0^2$, $\alpha $, and $d$ 
 are taken to be $\hat{Q}_0 = 0$, $\hat{\Omega}_0^2 = 0$, 
$\alpha = 0$, and $d = 0$
(for details, see Appendix 
\ref{app:arbitrary_functions}). 
As for  $\hat{\kappa}_0$, following \cite{Otani_Takahashi_Eriguchi_2009}, 
we take $\hat{\kappa}_0= 4.5$. 
Since there are no rotation and no meridional flows, 
the toroids are in stationary states
by the balance among the gravitational force of the central object, 
the Lorentz force and the gas pressure gradient.
If there could be very strong poloidal magnetic
fields near the central objects, stronger gravitational 
forces of the central objects could be balanced by
the {\it strong} magnetic forces near the central
objects. For such situations the toroids could be elongated
toward the central objects and have wider widths.

\cite{Fujisawa_Yoshida_Eriguchi_2012} showed that 
the poloidal magnetic field distributions substantially 
depend on the parameter $m$ in the arbitrary function $\mu(\Psi)$ 
for magnetized stars.  
In particular, they found that negative values of $m$ result in 
concentration of the poloidal magnetic fields near the symmetry axis of 
magnetized stars. Thus, it is expected that we may obtain toroids 
in which poloidal magnetic fields are concentrated near the inner 
edge of the toroids by choosing an appropriate value of $m$.

We show our numerical results for
critical configurations of $N = 1.5$ and $N = 3$
polytropic sequences in Table \ref{Tab:mu}
and density contours on the meridional cross section of $N=3$ polytropes
 with $m=-1.4$, $m=-0.7$,
$m=-0.1$ and $m=0.5$ in Fig. \ref{fig:tori}.
As seen from these table and figure, 
the critical distance 
decreases as the value of $m$ decreases.
The density distribution of the $m=-1.4$ 
toroid is stretched toward the central object 
because of the strong gravity
of the central object. In addition to this, 
the cusp-like shape at the inner edge of the toroids becomes 'sharper' as the value of
$m$ becomes smaller. We find the same tendency 
for the $N = 1.5$  polytropes.

The top panel of Fig. \ref{fig:tori2} shows the structures of the magnetic 
fields  on the meridional plane for the $N = 3$ critical configurations with 
$m=-1.4$ (solid curves) and $m=0.5$ (dashed curves). In this figure, 
the surfaces of the toroids are indicated by the tear-shaped closure curves. 
The structures of the magnetic fields for 
the model with $m=-1.4$ are remarkably different from 
those for the model with $m=0.5$. The shapes of 
the contours of the magnetic flux function for the toroid with
$m=0.5$ look nearly circle, but those for the toroid with 
$m=-1.4$ deform oblately.
Moreover, the magnetic field lines are more densely 
distributed near the inner edge region of the toroid  for
the toroid with $m=-1.4$ compared to those for the toroid with 
$m=0.5$. In other words,  
the magnetic fields are highly localized 
toward the central object for the toroid with $m=-1.4$ 
compared to those of the toroid with $m = 0.5$.

The panels in the middle left and the bottom left of Fig. \ref{fig:tori2}
show distributions of  $\log_{10} |\Vec{B}|$ on the equatorial plane, and 
the panels in the middle right and the bottom right of Fig. \ref{fig:tori2}
show values of each term in the right-hand side of  the first line of 
Eq. (\ref{Eq:first_int}) on the equatorial plane. Note that 
the horizontal axises of these figures range from 
$R_{inner}/r_e(=q_c)$ to $R_{outer} /r_e(=1)$.  
As seen from the middle left and the bottom left panels of Fig. \ref{fig:tori2}, 
the distributions of the magnetic fields are 
significantly different for the two equilibrium configurations. 
The ring of maximum magnetic field strength locates near the inner edge of the toroid  
for the model with $m=-1.4$, while it locates near the central region of the 
meridional cross section of the toroid for the model with $m = 0.5$. 
The toroids with $m = -1.4$ can sustain highly localized 
and strong magnetic fields in the nearer region from the 
central object compared to the toroids 
with $m = 0.5$ and can extend themselves toward 
the central object although the gravitational force
of the central object is much stronger there. 
This implies that the toroids with $m = -1.4$ can produce 
strong magnetic force near their inner edge, which is balanced against 
the gravitational force of the central compact object as shown later. 

In the middle right and the bottom right  panels of Fig. \ref{fig:tori2}, the gravitational 
potential of the central object, $-\hat{M}_c / 4\pi\hat{r}$
(thick dashed curve), the gravitational potential of 
the toroids, $\hat{\phi}_g$ (thin dashed curve), the magnetic 
potential (\citealt{Fujisawa_Yoshida_Eriguchi_2012}), 
$-\int \hat{\mu}\, d\hat{\Psi} - \hat{C}$ 
(dotted curve) and the sum of all the potentials (thick solid curve)
are shown as functions of $R/r_e$. 
Here,  physical quantities with 
$(\ \hat{} \ )$ are dimensionless quantities 
defined in Appendix \ref{app:dimensionless}. 
From these figures, it is clearly seen that
the magnetic force is the primary agent supporting the toroid against the 
gravitational force of the central object. 
The gradient of the gravitational potential of the 
central object for  the toroid with $m=-1.4$ is steeper 
than that for the toroid with $m=0.5$ 
because the $m=-1.4$ toroid is located closer to the 
central object than the $m=0.5$ toroid.
The magnetic potentials behave very differently for these 
two equilibrium configurations with different values of $m$. 
For the $m=0.5$
toroid, the magnetic potential curve has a substantial local minimum  
at $R / r_e \sim 0.9$.
For the $m=-1.4$ toroid, however, the magnetic potential 
curve is shallower and extends within a broader region 
and its slope is 
steepest near the inner edge of the toroid. 
As a result, the strong magnetic fields can exist 
near the inner edge region of the toroid and their magnetic force supports
the toroid against the gravitational force of the central 
compact object. In this way, the $m=-1.4$ toroid can be
in a stationary state even if the gravitational potential
becomes much steeper as approaching to the central object.

\subsection{Effects of the meridional flows
on the magnetized configurations}

\subsubsection{Basic features of magnetized configurations
with meridional flows}

\begin{figure*}
 \begin{center}
  \includegraphics[scale=0.7]{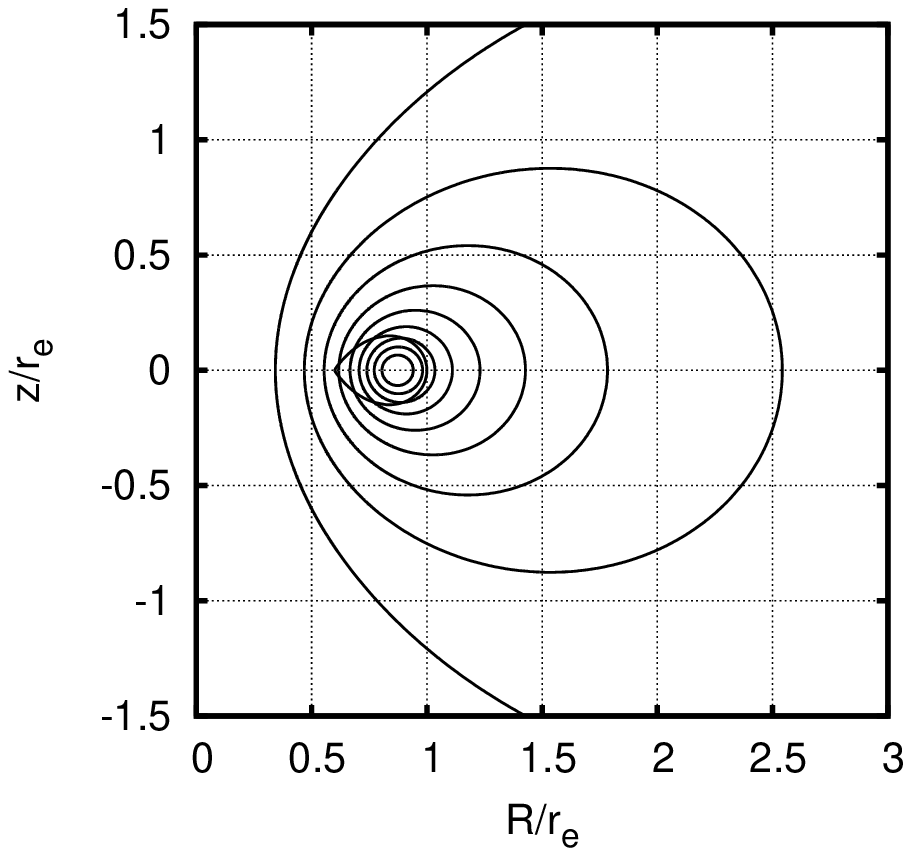}
  \includegraphics[scale=0.7]{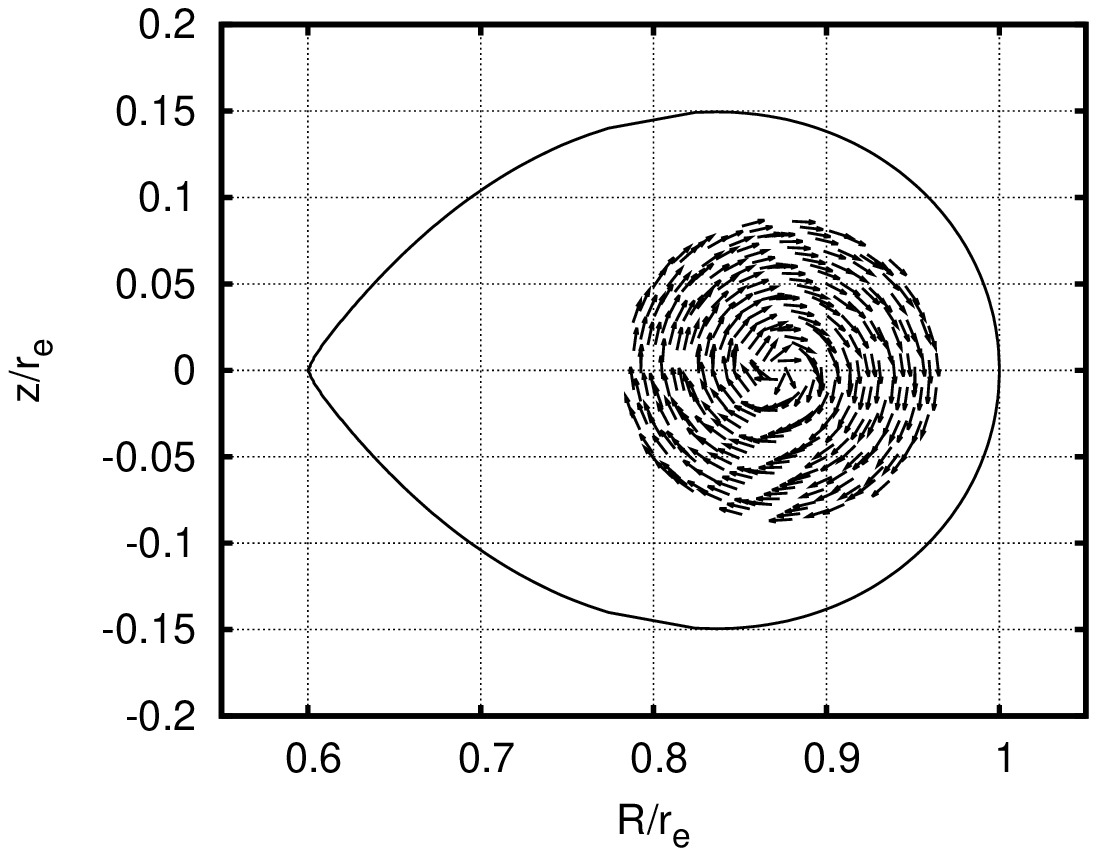}
  \includegraphics[scale=0.75]{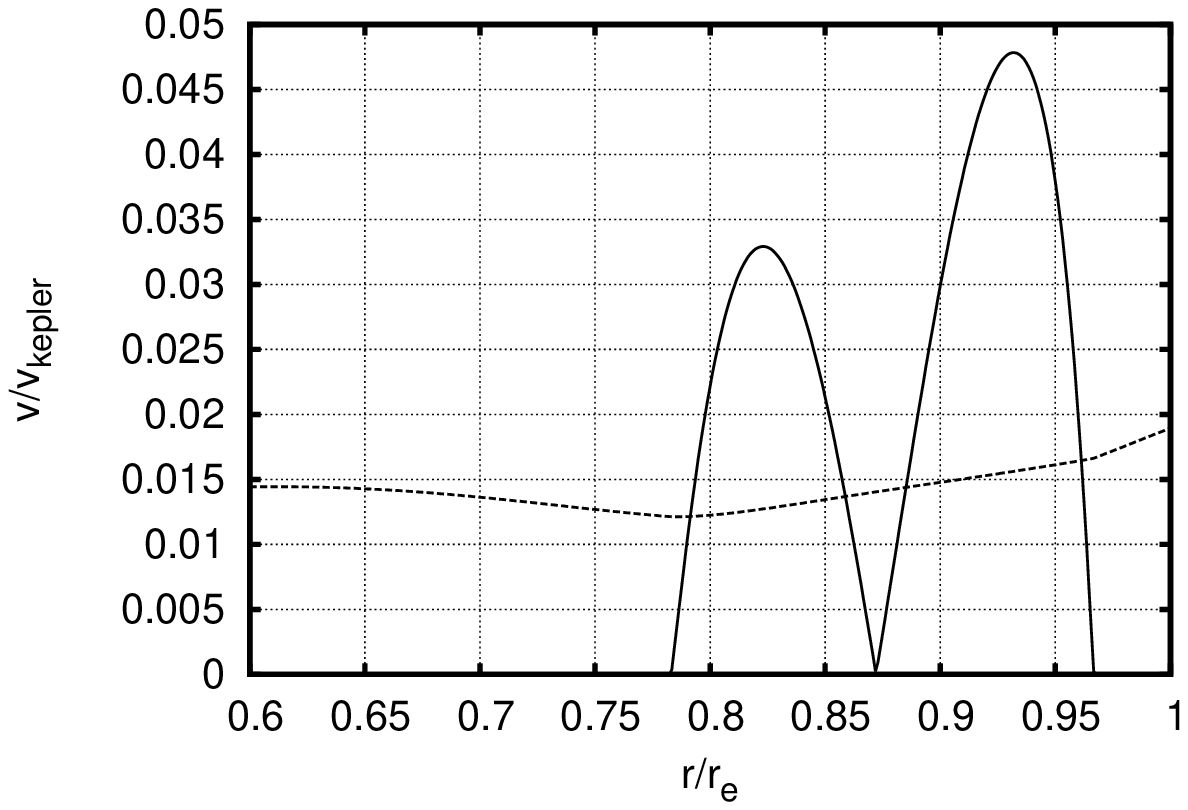}
 \end{center}
\caption{The poloidal magnetic fields (left), the meridional flow pattern (right) 
on the meridional plane for the toroid with $N=1.5$, $\hat{Q}_0 = 20$, $\hat{\Omega}_0 = 1.0 \times 10^{-5}$, 
$m=0$, $ q=0.6$, and $\hat{\kappa}_0 = 0.5$. 
The tear-shaped closure curves with 
a cusp-like structure indicate the surfaces of the toroids.
Length of vectors given in the right panel is not proportional to the speed of the meridional 
flows and does not have any physical meaning. 
The bottom panel shows the distribution of 
the absolute value of
the meridional velocity (solid line)
and the rotational velocity (dashed line) normalized by the local Kepler velocity 
on the equatorial plane. 
}
\label{Fig:basic_torus}
\end{figure*}

Concerning the parameters which appear in the arbitrary functions, 
to examine effects of the meridional flow on toroid structures, 
the following values are chosen in this subsection:
$\hat{\kappa}_0 = 0.5$, $\alpha = -0.5$, $d = 0.1$, 
$ m = 0$ and $\hat{\Omega}_0^2 = 1.0 \times 10^{-5}$. 
 Note that smaller values of $\hat{\kappa}_0$ 
result in more rapid meridional flows and that this small value 
of $\hat{\Omega}_0$ gives equilibrium configurations with almost no rotation.
Parameters for the rotation law are the same as those
in \cite{Yoshida_Yoshida_Eriguchi_2006}. 

The left panel of Fig. \ref{Fig:basic_torus} shows 
contours of the flux function $\Psi$ on the meridional plane for the critical configuration
of an $N = 1.5$ polytrope with meridional flows. 
Direction of the fluid velocity on the meridional cross section in the critical configuration
is shown in the right panel of Fig. \ref{Fig:basic_torus}.
Here, the lengths of the vectors are not proportional
to the absolute values of the fluid velocities. Note that the region 
where the meridional flows are present is only a part of 
the meridional cross section of the toroid 
because a particular functional form for $\hat{Q}'(\Psi)$ is used to avoid singular behaviors of 
the meridional flow 
near the toroid surface (see Appendix \ref{app:arbitrary_functions}). 
The bottom panel of Fig. \ref{Fig:basic_torus} displays the velocity distributions normalized 
by the local Kepler velocity, on the equatorial plane.
The solid and dashed curves denote the absolute value of 
the meridional velocity and the rotational velocity, respectively. 
As seen from this panel, the rotational velocity is
sub-Keplarian, because our rotational parameter $\hat{\Omega}_0$ is 
assumed to be small in this paper. 
On the other hand, the meridional velocity is slightly faster than the rotational velocity
in this parameter region ($\hat{Q}_0$ = 20).

As shown in Figs. \ref{fig:tori2} and \ref{Fig:basic_torus} 
(see, also, Table \ref{Tab:meridional_flow} given later), general structures of the toroids and 
their magnetic fields do not change 
significantly even when the fluid flows exist on the meridional plane. As argued later, 
however, the presence of the meridional flows
changes the density distributions of the toroids slightly 
and increases the critical distance a little bit.

\subsubsection{Critical distances for magnetized toroids
with meridional flows}
\label{Sec:3.2.2}
%
%
\begin{table*}
{\small
\begin{tabular}{rrrrrrrrrrrrrr}
  \hline 
  $\hat{Q}_0$ & $q_c$& $\hat{\mu}_0$ &   $r_e$(cm) & 
 $\mathfrak{M}/|W|$ & $\Pi/|W|$& $T/|W|$  &  $T_p / |W|$ &VC \\

\hline
&&&&\hspace{-20pt} $N=1.5$ \\
\hline

  0 & 0.597 & 3.047E+00 & 2.500E7  & 6.213E-01 &
  1.262E-01 & 8.322E-05 &  0.000E+00 & 2.656E-05 \\
 
  20 & 0.600 & 3.020E+00 & 2.514E7 & 6.226E-01 &
  1.255E-01 & 4.138E-04 & 3.185E-04 & 3.032E-05 \\
 
  40 & 0.605 & 2.942E+00 & 2.555E7 & 6.268E-01 &
  1.235E-01 & 1.431E-03 & 1.314E-03 & 4.271E-05 \\
 
  60 & 0.615 & 2.817E+00 & 2.627E7  & 6.338E-01 &
  1.199E-01 & 3.217E-03 & 3.063E-03 & 6.505E-05 \\

  80  & 0.629 & 2.653E+00 & 2.759E7 & 6.437E-01 &
  1.148E-01 & 6.101E-03 &  5.875E-03 & 1.595E-04 \\

  \hline
  &&&&\hspace{-20pt} $N=3$ \\
  \hline

   0 & 0.625 & 2.810E+00 & 3.551E7 & 6.583E-01 &
   1.133E-01 & 9.202E-04 & 0.000E+00 & 5.497E-05 \\
 
   20 & 0.628 & 2.784E+00 & 3.570E7 & 6.598E-01 &
   1.125E-01 & 1.464E-03 & 4.866E-04 & 3.642E-05 \\
 
   40 & 0.632 & 2.711E+00 & 3.622E7 & 6.639E-01 &
   1.101E-01 & 2.935E-03 & 1.842E-03 & 3.115E-05 \\

\hline
\end{tabular}
}
\caption{Physical quantities for the critical configurations with meridional flows. 
Model parameters are $\hat{\kappa}_0 = 0.5$, $\hat{\Omega}_0^2 = 1.0 \times 10^{-5}$,
and $ m = 0$. 
}
\label{Tab:meridional_flow}
\end{table*}
%
%
\begin{figure*}
\includegraphics[scale=0.7]{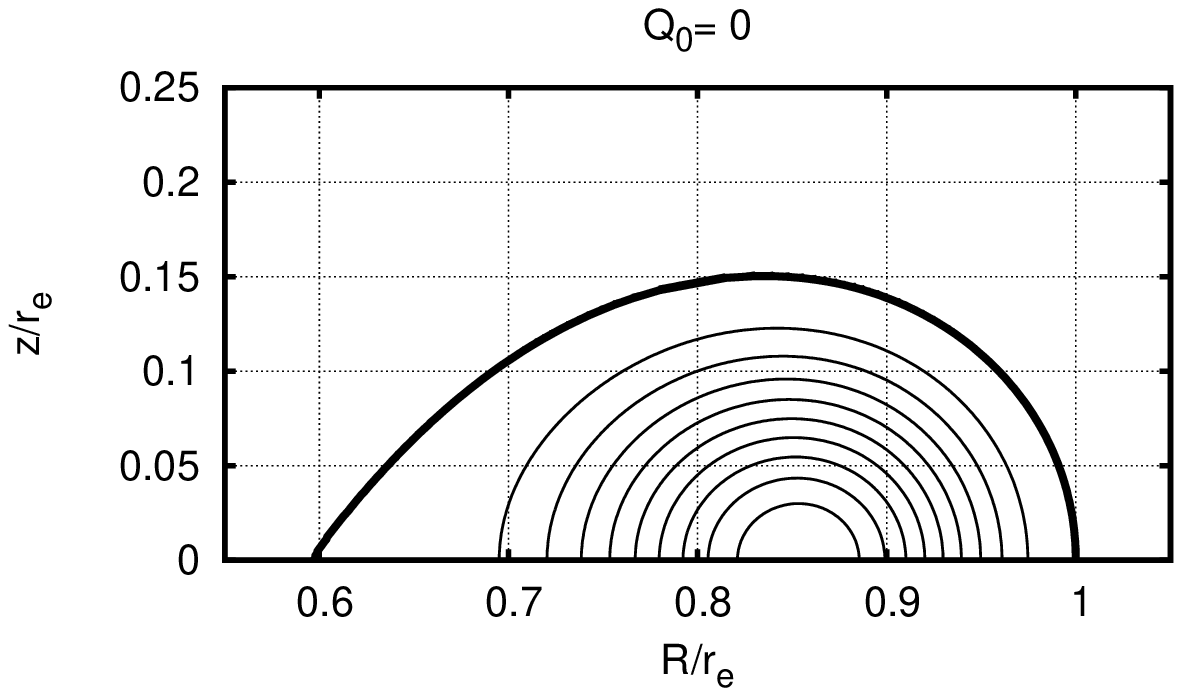}
\includegraphics[scale=0.85]{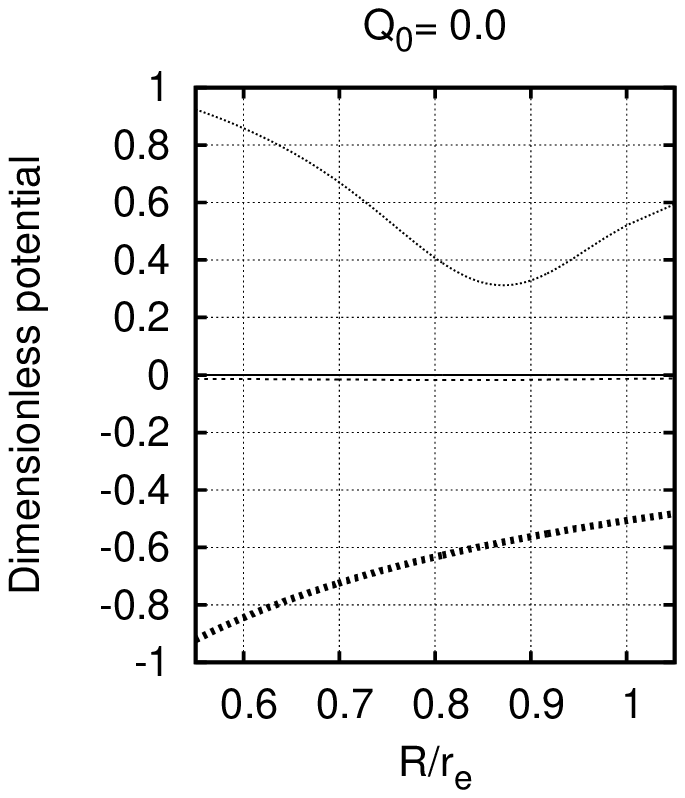}

\includegraphics[scale=0.7]{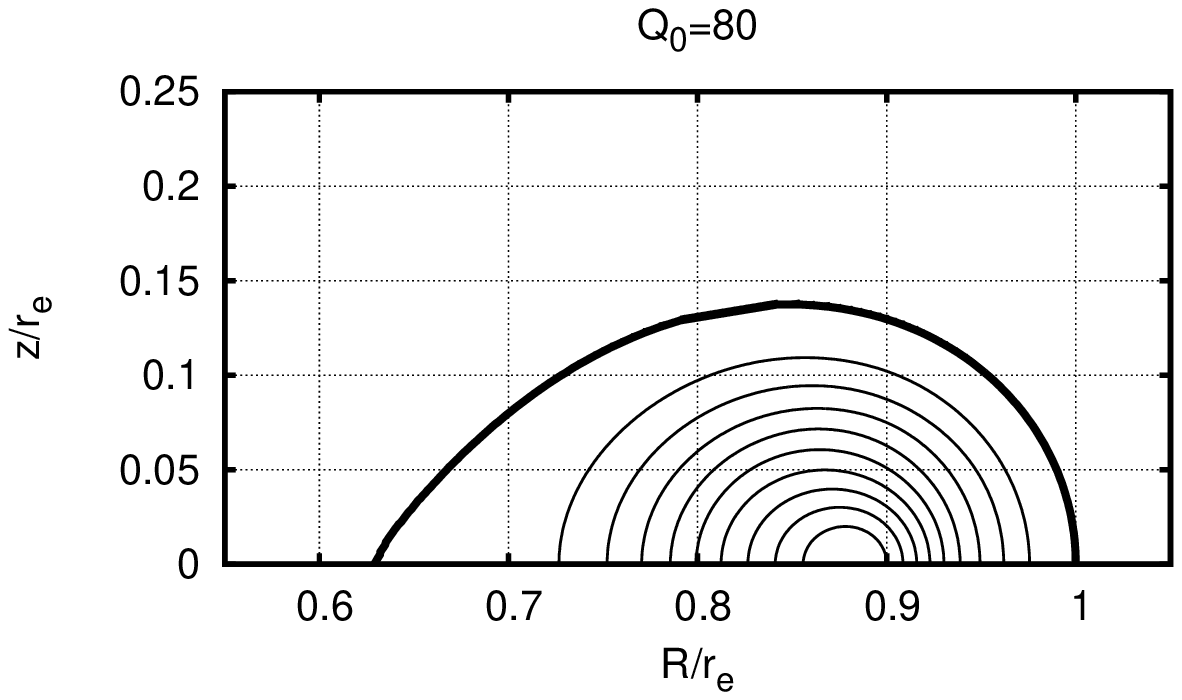}
\includegraphics[scale=0.85]{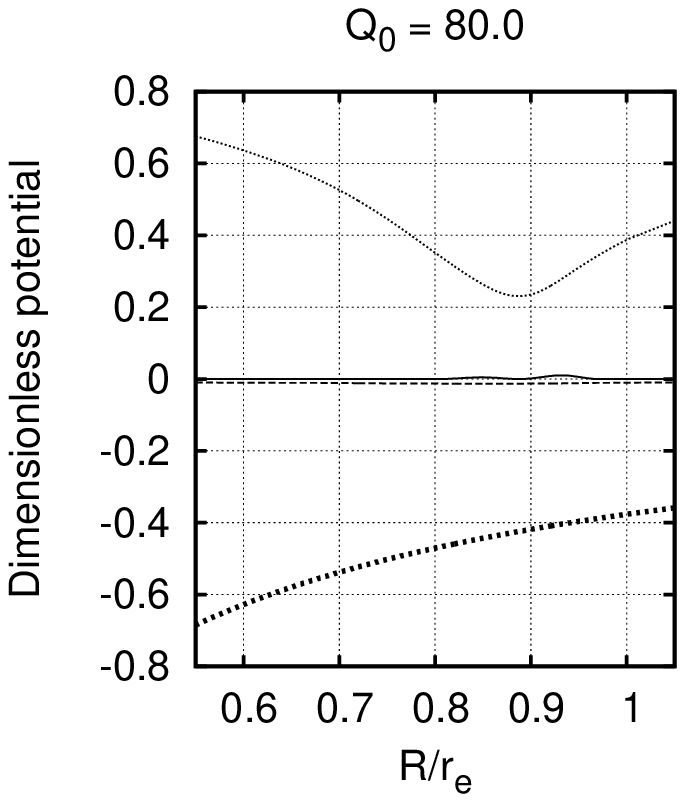}

\includegraphics[scale=0.6]{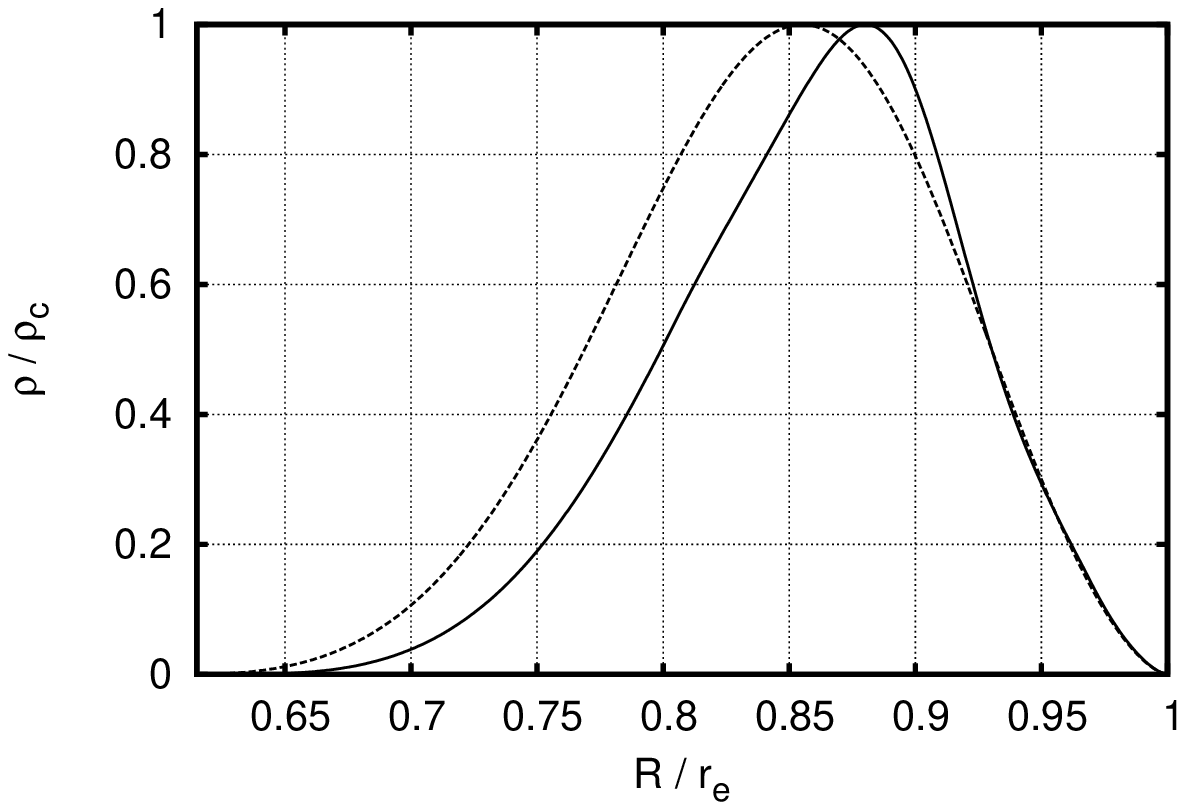}
\includegraphics[scale=0.6]{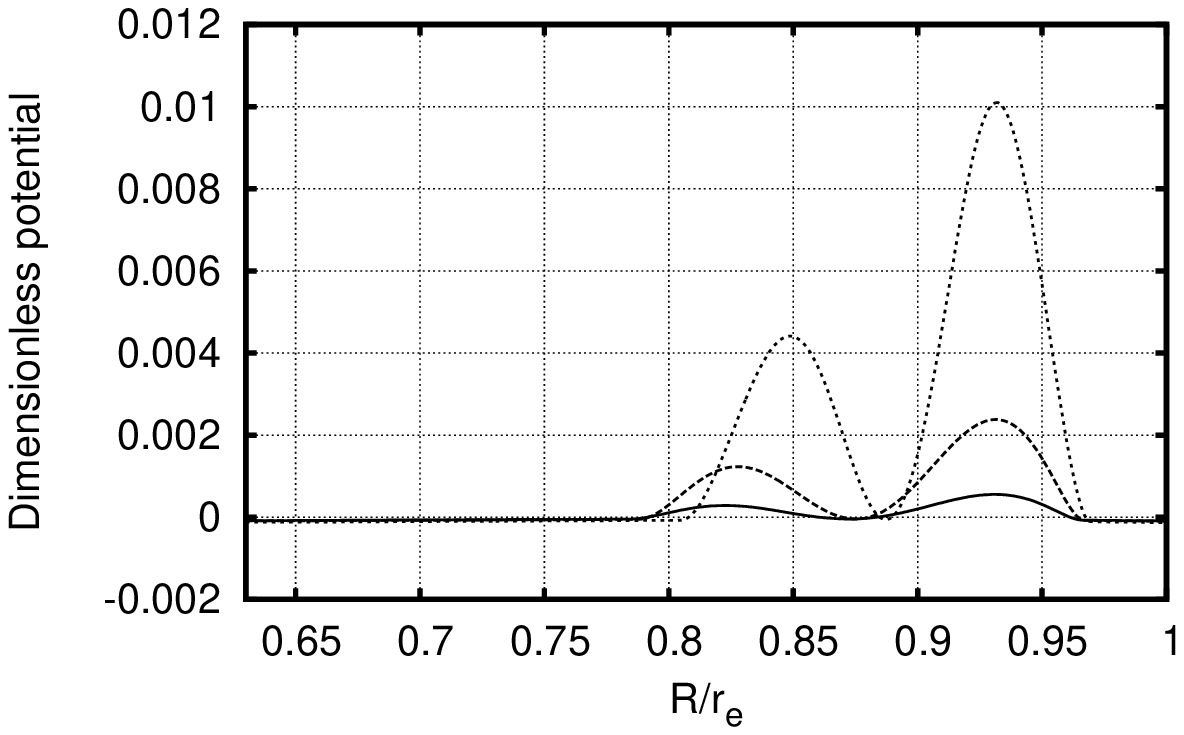}

\caption{Top left and middle left panels: 
Density contours on the meridional cross section for 
the critical configurations with $\hat{Q}_0=0$ (top left) and 
$\hat{Q}_0=80$ (middle left). The heavy curves indicate 
the surfaces of the toroids. The density difference between 
two adjacent contours is one-tenth of the maximum density.  
Top right and middle right panels: 
$\frac{1}{2}|\hat{\Vec{v}}^2| - \hat{R} \hat{v}_\varphi 
\hat{\Omega}$ (solid curve),  $-\hat{M}_c /4\pi \hat{r}$ 
(thick dashed curve), $- \int \hat{\mu} \, d \hat{\Psi} - \hat{C}$ 
(dotted curve) for the critical configurations with 
$\hat{Q}_0=0$ (top right) and $\hat{Q}_0=80$ (middle right), 
given as functions of $R/r_e$. 
The horizontal axis ranges from $R_{inner}/r_e(=q_c)$ to $R_{outer} /r_e(=1)$. 
Bottom left panel: 
Densities normalized by $\rho_c$ on the equatorial plane for the critical configurations 
with $\hat{Q}_0 = 0$ (dashed curve) and 
$\hat{Q}_0 = 80$ (solid curve), given as functions of $R/r_e$. 
The horizontal axises range from $R_{inner}/r_e(=q_c)$ to $R_{outer} /r_e(=1)$. 
Bottom right panel:  
The dimensionless velocity potential term,  
$\frac{1}{2}|\hat{\Vec{v}}^2| - \hat{R} 
\Vec{v}_\varphi \hat{\Omega}$, on the 
equatorial plane for the the critical configurations 
with $\hat{Q}_0 = 20$ (solid curve), $\hat{Q}_0 = 40$ (dashed curve) and 
$\hat{Q}_0 = 80$ (dotted curve), given as functions of $R/r_e$. 
The horizontal axis ranges from $R_{inner}/r_e(=q_c)$ to $R_{outer} /r_e(=1)$. 
The model parameters are $m = 0$,  $N=1.5$, $\hat{\kappa}_0 = 0.5$, 
and $\hat{\Omega}_0^2 = 1.0 \times 10^{-5}$.
}
\label{Fig:contours}
\end{figure*}
%
%

In order to study the influence of the 
meridional flows on the 
critical distances,  we calculate two polytropic sequences ($N=1.5$, $N=3$) by 
changing the value of $\hat{Q}_0$ for the $m=0$ toroidal current function. 
We fix the other parameters as $\hat{\kappa}_0 = 0.5$, $\alpha = -0.5$, $d = 0.1$, 
and  $\hat{\Omega}_0^2 = 1.0 \times 10^{-5}$ in this subsection.
Physical quantities for several models belonging to the two  polytropic sequences 
are tabulated in Table \ref{Tab:meridional_flow}. 
Here, $T_p$ denotes the kinetic energy of the 
meridional flow.

As seen from this table, the energy ratio $T_p / |W|$ is much smaller 
than that of ${\cal M} / |W|$, thus, the 
kinetic energy of the meridional flow 
is much smaller than the magnetic energy for the models obtained in the present study.
This means that the magnetic forces mainly support the toroids against 
the gravitational forces of the central objects 
even when the meridional flow becomes stronger  in the present parameter space. 
The meridional flow cannot change global structures  
of the toroids and their magnetic fields significantly. 
However,  the energy ratio $T_p / |W|$ reaches $\sim 1.0 \times 10^{-3}$
when $\hat{Q}_0 > 40$ $(N=1.5)$ and $\hat{Q}_0 = 40 $ $(N=3)$.
In fact, the critical distance $q_c$ increases 
as $\hat{Q}_0$ increases as shown in Table \ref{Tab:meridional_flow}. This implies that 
the toroids tend  to 
shed their mass when the meridional flow becomes stronger. 
In some sense, therefore, the influence 
of the meridional flow on the density distribution 
of the toroids should not be
considered to be small.
The rotational velocities of these models are
 also sub-Keplarian. The typical value is about 2 \% of the Kepler velocity
which is similar to that given in Fig.\ref{Fig:basic_torus}.
On the other hand,  the meridional flow velocity 
is about several times as large as the rotational 
velocity. The maximum velocity of the meridional flow for the
$\hat{Q}_0 = 80$ model reaches  about 10 times as large as 
that of the rotational velocity. 

In order to clarify the effects of the 
meridional flows, 
let us investigate the density distributions of toroids and 
the profiles of potential terms on the
equatorial planes for the $N=1.5$ toroids  with and without meridional flows. 
Fig. \ref{Fig:contours} shows the density contours on the meridional plane 
(top left and middle left) and the profiles of the potentials
on the equator (top right and middle right). 
Bottom  panels of Fig. \ref{Fig:contours} show 
the profiles of  the density on the equator
 (left) and the velocity potential term,  
$\frac{1}{2} |\hat{\Vec{v}}^2| - \hat{R}
 \hat{v}_\varphi \hat{\Omega}$, on the equator (right) as functions of $R/r_e$.
In the top right panel of Fig. \ref{Fig:contours}, each curve denotes 
$\frac{1}{2}|\hat{\Vec{v}}^2| - \hat{R} \hat{v}_\varphi 
\hat{\Omega}$ (solid curve),  
$-\hat{M}_c /4\pi \hat{r}$ (thick dashed curve), $\hat{\phi}_g$ (thin dashed curve) and 
$- \int \hat{\mu} d \hat{\Psi} - \hat{C}$ (doted curve).
As seen from these profiles, in both the models, the gravitational potentials of the toroid make a tiny 
contribution to the equilibrium solutions and  the balance between 
the term $-\int \hat{\mu} d \hat{\Psi} - C $ and the gravitational potential of the central object mainly determines 
the stationary states of the toroid. 
Taking a detailed look at the middle right panel of Fig. \ref{Fig:contours}, 
we observe that the potential terms due to the meridional flow and rotation 
(solid curve) show their maximum value near the radius of  $R/r_e = 0.95$  
for the $N = 1.5$ models
with $\hat{Q}_0 = 80.0$. This very tiny protuberance is considered to appear due to 
the presence of the meridional flows because as shown in Table \ref{Tab:meridional_flow}, 
the kinetic energy due to rotation is negligible small.
More detailed structures  of the velocity potential terms can be seen 
by enlarging these tiny protuberances.
The bottom right panel of Fig. \ref{Fig:contours} shows the 
profiles of $\frac{1}{2}|\hat{\Vec{v}}^2| - \hat{R} \hat{v}_\varphi \hat{\Omega}$ on the equator
for $N = 1.5$ polytopes 
with $\hat{Q}_0 = 20$ (solid curve), 
$\hat{Q}_0 = 40$ (dashed curve) and
$\hat{Q}_0 = 80$ (dotted curve). 
These profiles have double peaks which locate 
at $\hat{R} \sim 0.85$ and $\hat{R} \sim 0.92$.
These double peaks appear from the balance 
of the density distributions of 
the toroids and the meridional flows.
As we have described in Sec. \ref{Sec:intro}, the presence of 
the poloidal velocity fields results in reducing the density of 
toroids (see Eq. (\ref{eq:first_integral})). For our numerical examples,  the presence of the poloidal velocity fields 
decreases the density on the equatorial plane around radii of $\hat{R} \sim 0.85$ and 
$\hat{R} \sim 0.92$, which can be observed in the bottom two panels of Fig.\ref{Fig:contours} 
(more detailed considerations are given below). 

The top left and middle left panels of Fig. \ref{Fig:contours} 
show the density distributions of  $N=1.5$ toroids with
$\hat{Q}_0 = 0.0$  and $\hat{Q}_0 = 80$, respectively.
Comparing the top left panel with the middle left panel, we see that 
the matter distribution of the toroid 
with $\hat{Q}_0 = 80$ is shifted  outward 
slightly. 
The bottom left panel of Fig.\ref{Fig:contours} 
displays the density profiles on the equator 
for each toroid. 
The dotted and solid curves  denote the 
density profiles for the models 
with $\hat{Q}_0 = 0.0$ 
and $\hat{Q}_0 = 80$, respectively. As seen 
form this panel, the meridional flows shift 
the place where the density takes its maximum
value  outward and make 
the density gradient around $\hat{R} \sim 0.92$ 
steeper.
This is due to the double peak structure 
of the velocity potential profiles.
The inner peak of this potential affects to 
decrease the density around $\hat{R} \sim 0.85$ 
where the density takes the maximum value 
if there is no meridional flow.
On the other hand, the outer peak of 
this potential also leads to decrease in 
the density around $\hat{R} \sim 0.92$. 
Since the density decreases around 
$\hat{R} \sim 0.85$ and $\hat{R} \sim 0.92$ 
by the presence of the meridional flows,  
the place where the density becomes  
maximum moves outward and the density 
gradient becomes steeper if the toroids 
have rapid meridional flows 
($\hat{Q}_0 = 80$ model).
These effects result in decreasing the 
critical distances.

Next, we deal with the influence of the 
equation of state 
on the critical distance. As we have seen in Table \ref{Tab:meridional_flow},
the critical distances of the $N=3$ toroids are larger than those
 of the $N=1.5$ toroids.  The same tendency 
in the equilibrium configurations without meridional flows found in \cite{Otani_Takahashi_Eriguchi_2009}. 
This is because that the mass shedding from the inner edge of the toroids is more
likely to occur for softer equations of state. 

\subsubsection{Effects of meridional flows on equilibrium configurations 
with highly localized poloidal magnetic fields}

\begin{figure*}

\includegraphics[scale=0.6]{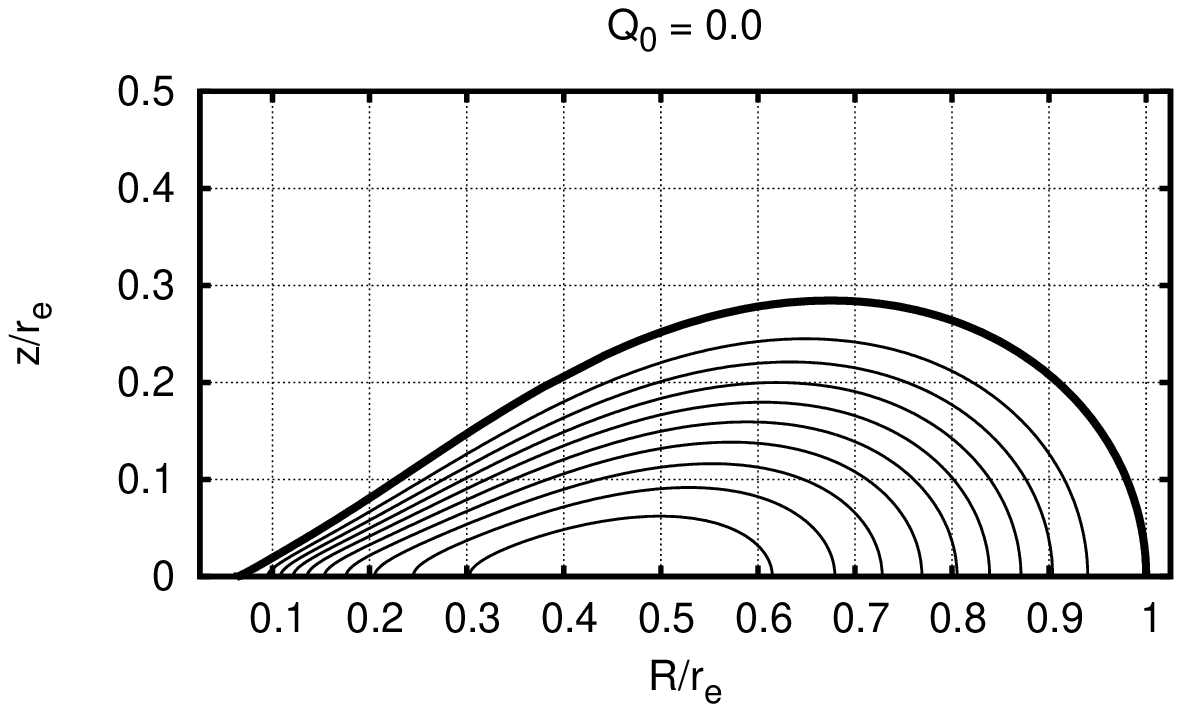}
\includegraphics[scale=0.6]{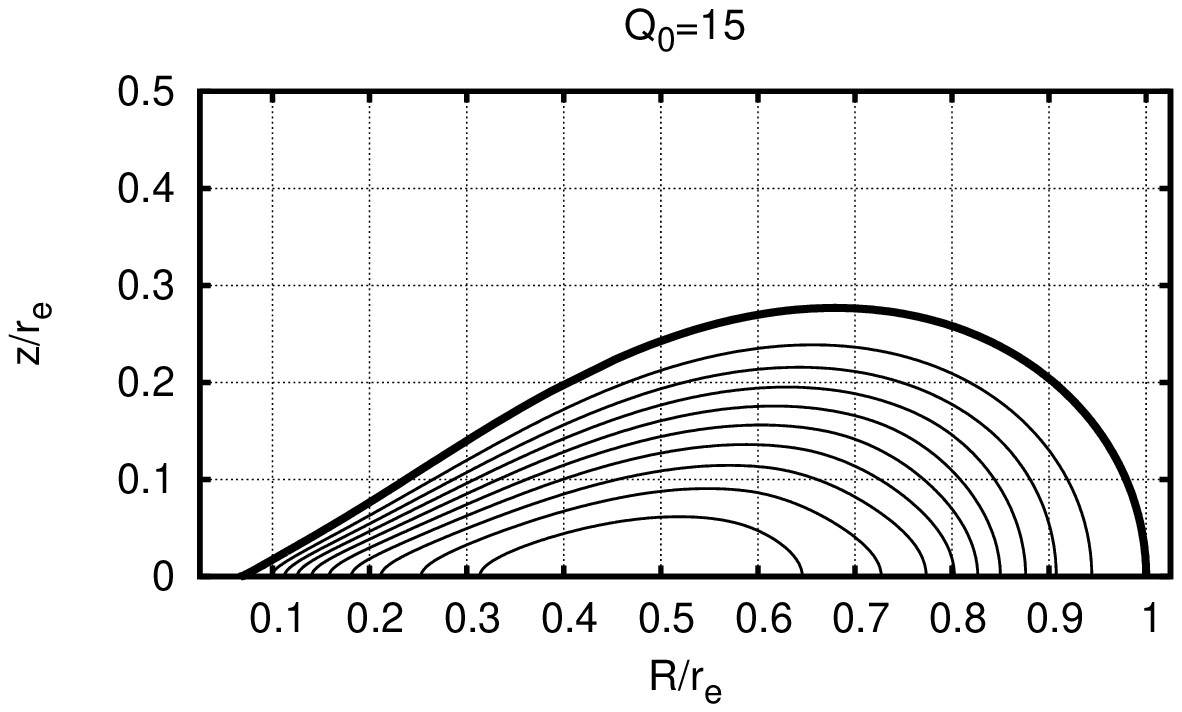}

 \includegraphics[scale=0.65]{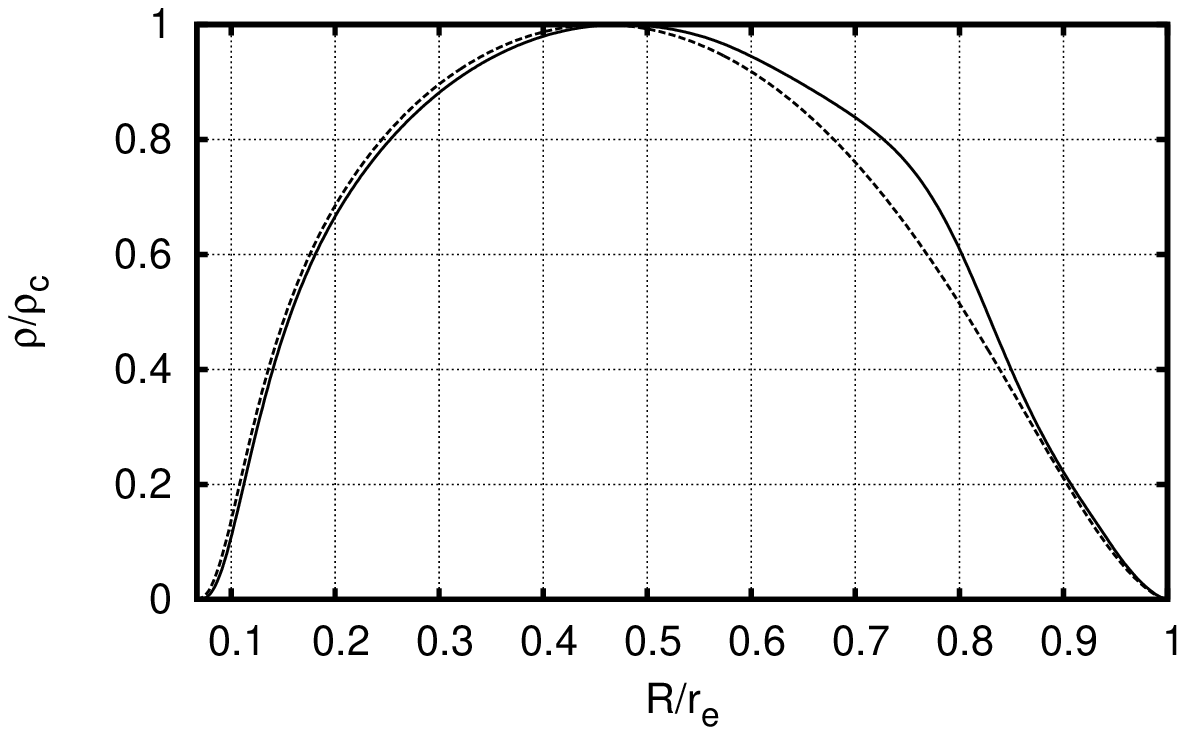}
\includegraphics[scale=0.6]{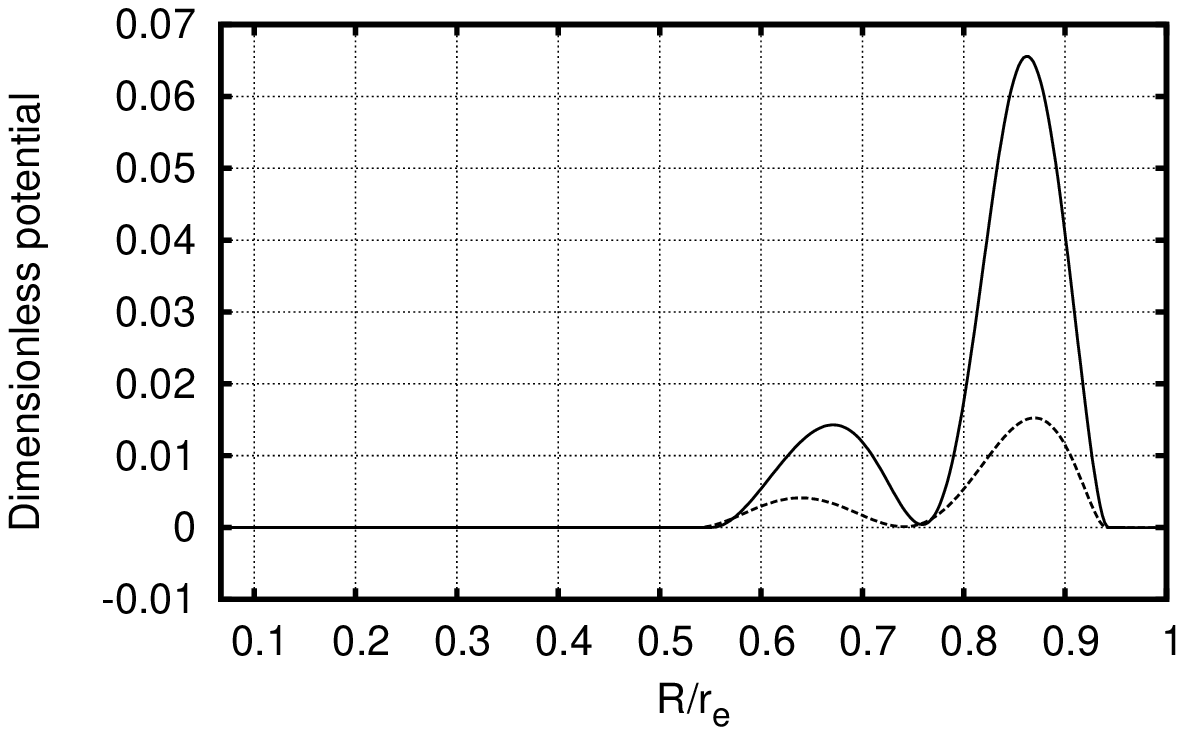}
 \caption{
Top panels: The density contours  on the meridional cross section for 
the critical configurations with $\hat{Q}_0 = 0.0$ (left panel) and $\hat{Q}_0 = 15$ (right panel).  
The heavy curves denote the surfaces of the toroids. The density difference between 
two adjacent contours is one-tenth of the maximum density. 
Bottom left panel: Densities normalized by $\rho_c$ on the equatorial plane 
for critical configurations with $\hat{Q}_0 = 0 $ (dashed curve) and $\hat{Q}_0 = 15$ (solid curve), 
given as functions of $R/r_e$.  The horizontal axis ranges from $R_{inner}/r_e(=q_c)$ to $R_{outer} /r_e(=1)$. 
Bottom right panel : The dimensionless velocity potential term,  
$\frac{1}{2}|\hat{\Vec{v}}^2| - \hat{R} \hat{\Vec{v}}_\varphi \hat{\Omega}$, on the
equatorial plane for critical configurations 
with $\hat{Q}_0 = 10$ (dashed curve), $\hat{Q}_0 = 15$ (solid curve), given as functions of $R/r_e$. 
The horizontal axis ranges from $R_{inner}/r_e(=q_c)$ to $R_{outer} /r_e(=1)$. 
The model parameters are $m = -1.4$,  $N=1.5$, $\hat{\kappa}_0 = 0.5$, 
and $\hat{\Omega}_0^2 = 1.0 \times 10^{-5}$.
}
\label{fig:profiles}
\end{figure*}

\begin{figure*}
 \includegraphics[scale=0.6]{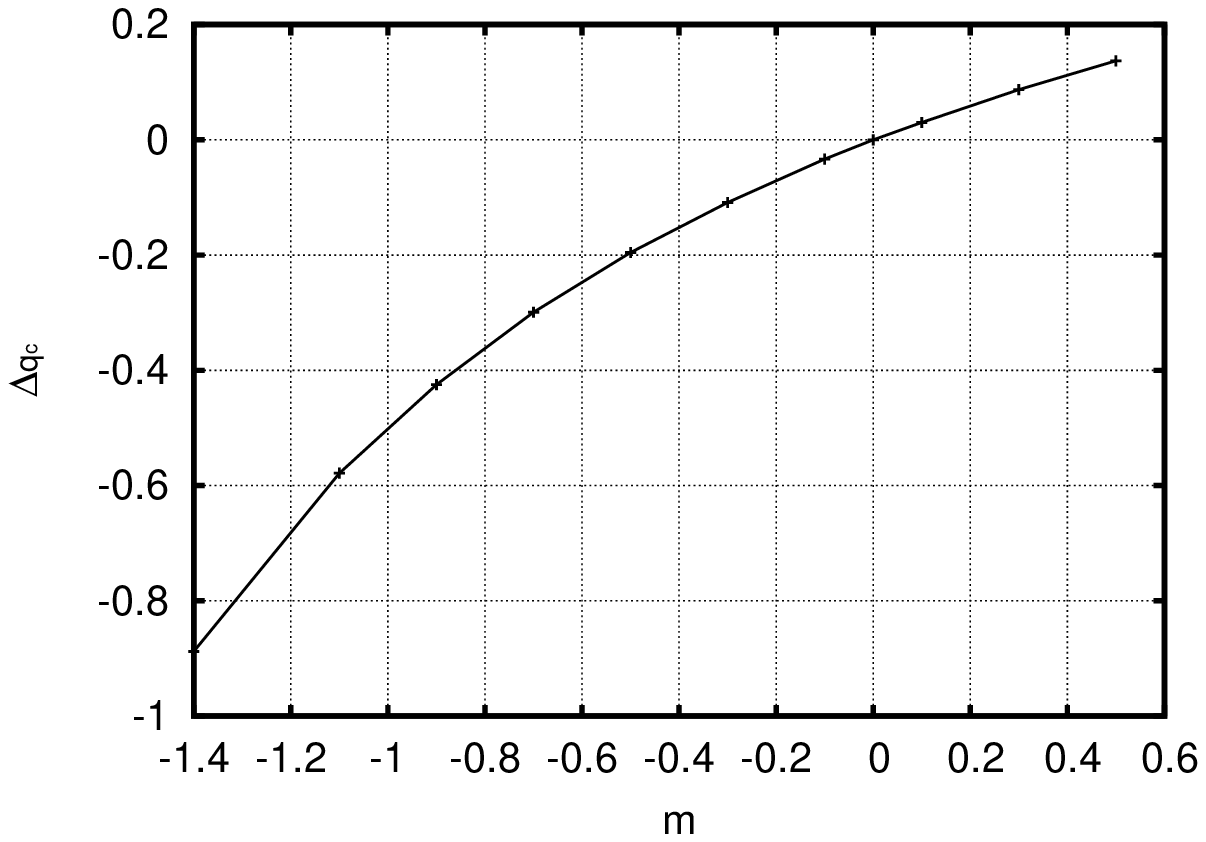}
 \includegraphics[scale=0.6]{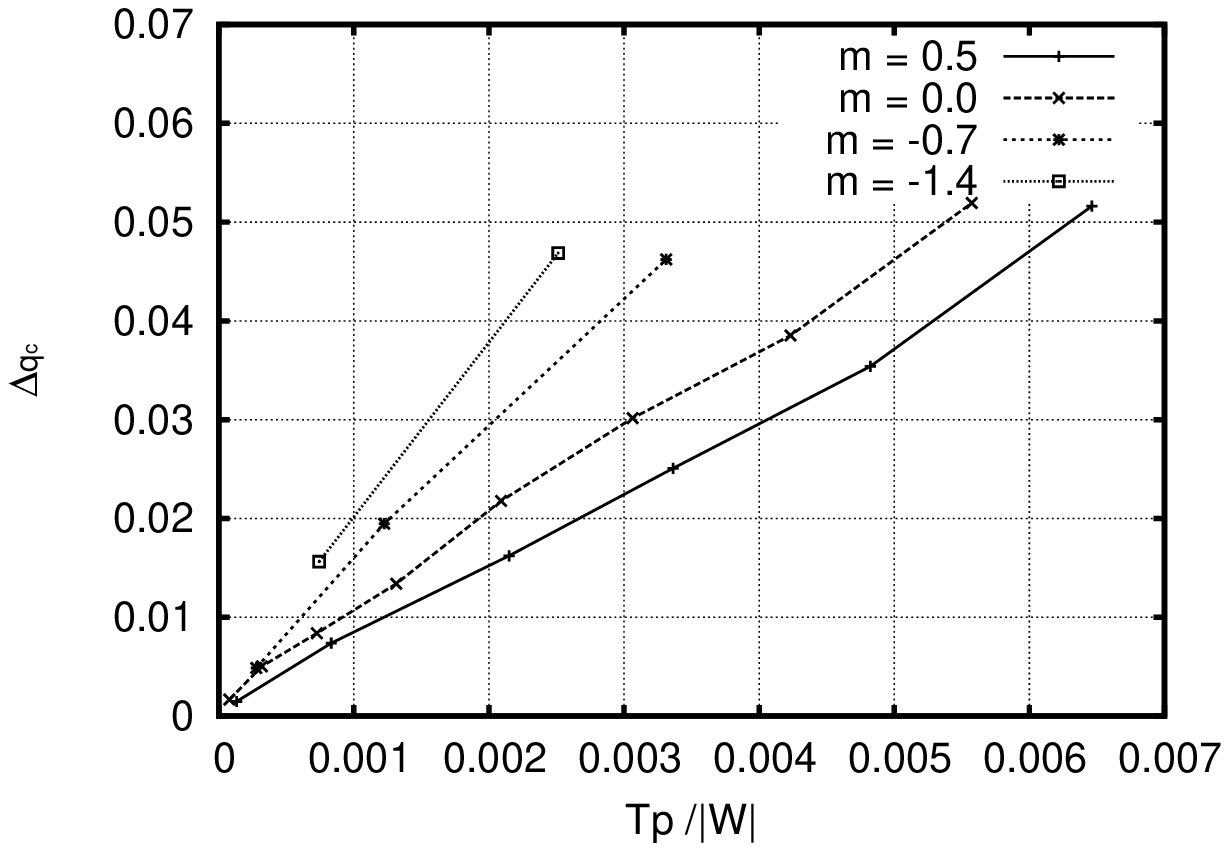}
 \caption{$\Delta_{q_c}(0,m,;0,0)$ vs. $m$ (left panel) and 
 $\Delta_{q_c}(\hat{Q}_0,m,;0,m)$ vs. $T_p /|W|$  for the critical 
 configurations with $m=0.5,\ 0,\ -0.7,\ -1.4$ (right panel).  
Here, $\Delta_{q_c}(Q_1, m_1;Q_2, m_2) \equiv q_c(Q_1, m_1)/q_c (Q_2, m_2) 
- 1$ and $T_p/|W|$ is given as a function of $\hat{Q}_0$ for the sequence of 
the critical configurations. }
\label{fig:dq}
\end{figure*}

Finally, we unveil the effects of the meridional flows on structures of 
the toroid having highly localized poloidal magnetic fields.
We consider the $N=1.5$ toroid models only  in this subsection because 
basic properties are independent of the equation of state. 
Fig. \ref{fig:profiles} displays typical models characterized by  
highly localized magnetic fields with and without strong meridional flows. 
Here, we take $m=-1.4$, with which highly localized poloidal magnetic 
fields are obtained inside the toroid as  argued in Sec. \ref{Sec:numerical_results}. 
Other parameters are 
$\hat{\kappa}_0 = 0.5$, $\alpha=-0.5$, $d=0.1$, and $\hat{\Omega}_0^2 = 1.0 \times 10^{-5}$, 
which are the same as those used in Sec. \ref{Sec:3.2.2}. 
The top two panels of Fig. \ref{fig:profiles} show the density distributions on the meridional  cross section 
for the toroids with no meridional flow (left) and with strong meridional flows (right). 
The bottom two panels  of Fig. \ref{fig:profiles} display the densities (left) and the velocity potential terms (right)
on the equatorial plane as functions of $R/r_e$. In the bottom left (right) panel, the solid and 
dashed curves are correspond to the models with $\hat{Q}_0=15$ ($\hat{Q}_0 = 15$) 
and $\hat{Q}_0 = 0$ ($\hat{Q}_0 = 10$), respectively. 
Comparing the bottom two panels of Fig. \ref{fig:profiles} to those of Fig. \ref{Fig:contours}, we observe that  
the density profiles of the $m=-1.4$ models are substantially different from those of 
the $m=0$ models though behaviors of their velocity potential terms are 
similar in the sense that they show similar double peaks. 
The toroids with highly localized magnetic fields (the $m=-1.4$ models) 
are extended inward due to the strong gravity of  the central object 
in comparison with the $m=0$ models (Compare the bottom left panels 
of Figs. \ref{Fig:contours} and \ref{fig:profiles} ). 
As shown in the bottom left panel of Fig. \ref{fig:profiles}, 
the positions of the maximum density rings 
for the $m=-1.4$ models are nearly independent of values of  $\hat{Q}_0$,  
but the density gradient of the $\hat{Q}_0=15$ model is steeper than that of 
the $\hat{Q}_0=0$ model around $\hat{R} \sim 0.85$ because of  the presence of 
the meridional flows.  
As a result, the matter distribution around  
$\hat{R} \sim 0.7$ is stretched outward.
The presence of the meridional flows also decreases the critical distance slightly 
as shown in the bottom left panel of Fig. \ref{fig:profiles}, which can be seen more clearly 
for the $m=0$ models (see the bottom left panel of Fig. \ref{Fig:contours}). 

As we have exhibited in numerical examples so far, values of $q_c$ 
decrease as values of $m$ decrease, while they increase as 
values of  $\hat{Q}_0 (>0) $ increase. 
In other words,  the poloidal magnetic fields generated by positive toroidal currents 
are apt to expand the toroids to the directions normal to the equi-flux function surfaces 
in particular when the magnetic fields are highly localized around the inner edge of the toroids 
and the meridional flows act as an agent for shrinking the region where the fluid matter occupies.  
In order to quantify the influence of the highly localized 
magnetic fields and the meridional flows on the critical distance, 
we introduce a quantity $\Delta_{q_c} (Q_1,m_1; Q_2, m_2)$, defined by
\begin{eqnarray}
\Delta_{q_c}(Q_1,m_1; Q_2, m_2) \equiv 
 \frac{q_c(Q_1, m_1)}{q_c (Q_2, m_2)} - 1 \,, 
\end{eqnarray}
where $q_c(Q_0, m_0)$ is the critical distance of the equilibrium sequence of the toroid 
characterized by $N=1.5$, $M_t/M_c=2.0\times 10^{-2}$,  $\hat{Q}=Q_0$, and $m=m_0$. 
Positive (Negative) values of $\Delta_{q_c}$ mean that the critical distance for the sequence 
with $(Q_1, m_1)$ increases (decreases) or its width of the toroid decreases (increases) 
compared to that with $(Q_2, m_2)$. 
In the left and right panels of the Fig. \ref{fig:dq}, $\Delta_{q_c}(0,m; 0,0)$ is given 
as a function of $m$ and  $\Delta_{q_c}(\hat{Q_0},m; 0., m)$'s are given as 
functions of  $T_p/|W|$ for several fixed values of $m$, respectively. 
The left panel shows that the values of 
$\Delta_{q_c}$ range from about 
$-0.9$ to $0.2$. 
Highly localized magnetic fields (for models with
negative values of  $m$) 
show the significant influence on the critical 
distances. On the other hand, the right panel 
shows that  $\Delta_{q_c}$ can reach about 
$0.05$ due to the effects of the meridional flows.
Regardless of the sign of $m$, the values of 
$T_p / |W|$ range up to $\sim 0.006$ and the
maximum value of $\Delta_{q_c}$ can reach 0.05 as the
values of $\hat{Q}_0$ are changed. Thus, the 
maximum value of $\Delta_{q_c}$  
 would be 0.05  when the meridional flows exist. 
This means that the influence of 
the meridional flows on the critical distances 
is much smaller than that of the highly localized magnetic
fields, but it is certainly true that the meridional flows work as an increasing
factor for the critical distances. We also find that 
the effects of the poloidal magnetic fields and the meridional flows may nearly cancel out 
for the toroids characterized by $T_p / |W| \sim 0.005$ and $m = -0.2$.
For this model, the critical distance or the width of the toroids is similar to that of the model 
with $T_p / |W| =0$ and $m = 0$. As expected in Sec. \ref{Sec:1.1}, thus, we confirm that the oppositely working effects 
of the highly localized magnetic fields and the meridional flows result in nearly no change 
in the critical distance for some particular toroid model. 

\section{Discussion and concluding remarks}

\subsection{Strength of the magnetic fields inside the toroids}

\begin{figure*}
 \begin{center}
\includegraphics[scale=0.65]{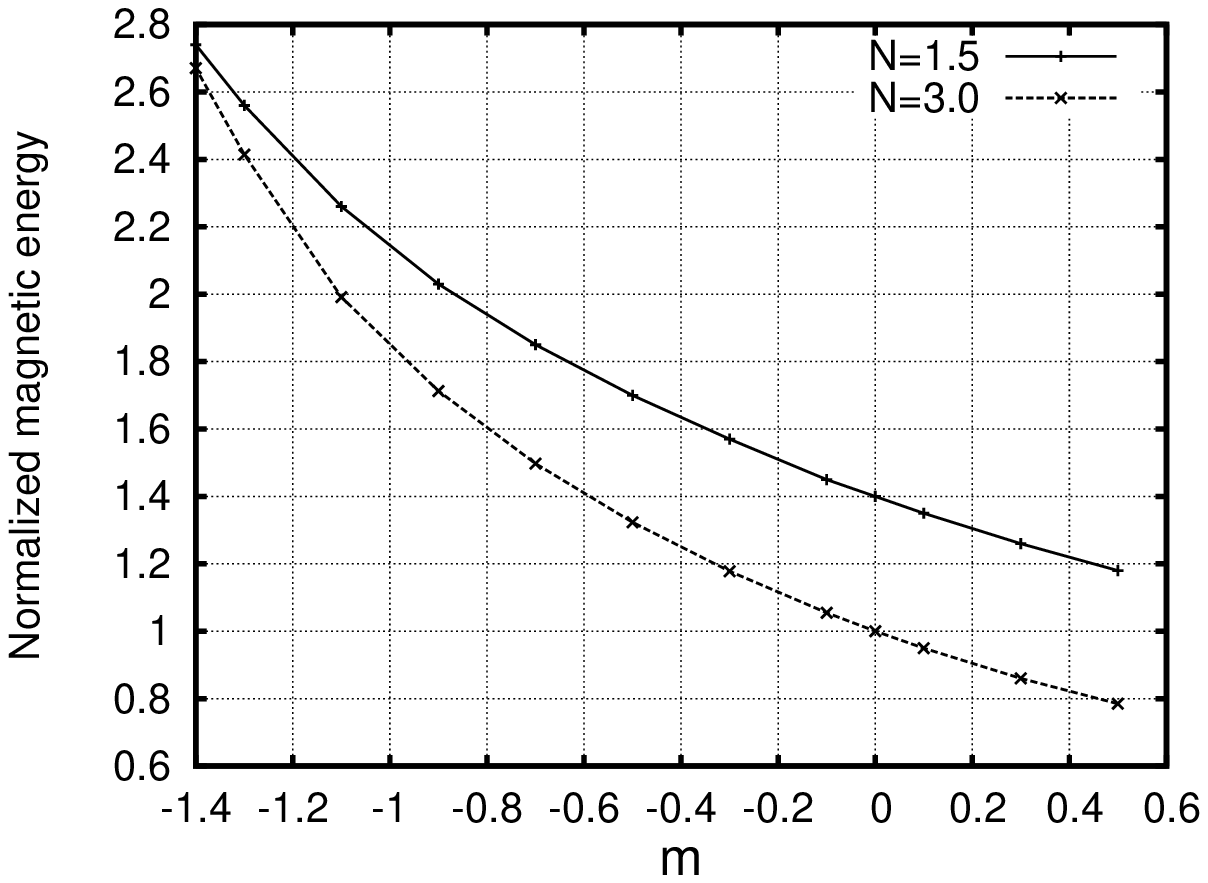}
 \end{center}
 \caption{
The magnetic energy normalized by the value of 
the equilibrium configuration with $m=0.0$ and  $N=3$, given as 
functions of $m$. The solid and dashed curves correspond to 
the $N=1.5$ and $N=3$ polytropes, respectively.
}
\label{fig:B_e}
\end{figure*}

As shown in the middle left and bottom left panels of Fig. \ref{fig:tori2}, 
for some particular set of the parameters, 
the toroids in the critical states have very strong magnetic fields and their strength are 
about $10^{15}$ G if we take $M_c = 5.0  M_\odot$ and 
$\rho_{\rm{c}} = 1.0 \times 10^{11} \rm{g/cm^3}$. 
Fig. \ref{fig:B_e} shows the magnetic energy of the toroids as functions of 
$m$. Here, the magnetic energy is normalized by that of the $m = 0.0$ and $N=3$ model.
From this figure, it is found that the magnetic energy 
becomes larger as the value of $m$ decreases.  
The magnetic energies of the $m=-1.4$ models are nearly
three times larger than that of the $m=0.0$ and $N=3$ model.  
Therefore, we conclude that larger magnetic energy 
can be sustained in the toroids whose magnetic fields are highly localized around the 
inner edge of the toroids and which locate closer to the central compact object.


\subsection{Critical distances for magnetized toroids with 
meridional flows and highly localized magnetic fields}

Assuming the toroidal current function $\mu$ to be constant ($m=0$), 
\cite{Otani_Takahashi_Eriguchi_2009} showed that there 
appears a critical distance in the self-gravitating toroids
with the magnetic fields and  that the critical distances are  much 
larger than the radii of the inner-most stable circular
orbit (ISCO) of the Schwarzschild black hole with the mass $M=M_c$. 
In this study, we show that their conclusions hold true when the meridional 
flows are taken into account and the functional form of $\mu$ is generalized 
to the cases of $m \geq 0$. 
For the cases of $m < 0$, however, we find that  the critical distance can be much 
shorter than that of the $m = 0$ case. 
The radius of ISCO of the Schwarzschild black hole with the mass $M=M_c$, $r_\mathrm{ISCO}$, 
is given by  
\begin{eqnarray}
r_\mathrm{ISCO} = \frac{6 G M_c}{c^2} \, .
\end{eqnarray}
If $M_c = 5.0 M_\odot$, which is the fiducial value in this study,  
$ r_\mathrm{ISCO} \sim 4.43 \times 10^{6} $cm. 
For the  $N = 3$ models with $m=0.0$ and $m = -1.4$, using results given in Table \ref{Tab:mu}, 
we respectively obtain 
\begin{eqnarray}
 r_\mathrm{in} 
= q \times r_e  
 \sim 2.30 \times 10^7 \mathrm{cm}
 > r_\mathrm{ISCO}\,,
\end{eqnarray}
and 
\begin{eqnarray}
 r_\mathrm{in} 
 = q \times r_e 
 \sim 1.08 \times 10^6 \mathrm{cm}
 \lessapprox r_\mathrm{ISCO}\,.
\end{eqnarray}
Therefore, the critical distance can be the same order 
or even smaller than the radius of the ISCO
 in this parameter space.
Since this study is done within the framework of  Newtonian gravity, 
the quantitative evaluation is not correct if $r_\mathrm{in}=O(r_\mathrm{ISCO})$, 
while the results given in the present study are reasonable as long as $r_\mathrm{in} \gg r_\mathrm{ISCO}$. 
We need to use general relativity for the toroids with $r_\mathrm{in}=O(r_\mathrm{ISCO})$. 
However, this is beyond the scope of the present study as mentioned before. 

\subsection{Concluding remarks}

In this paper, we have investigated and calculated 
the stationary states of magnetized self-gravitating 
toroids with meridional flows and various kinds of the magnetic fields 
around central compact objects.
As a result, we have obtained the toroids with 
strong meridional flows and with strong poloidal magnetic fields. 
Our findings and conjectures are summarized as follows.

\begin{enumerate}
 \item Choosing the functional forms of arbitrary functions, 
       we can change the strengths of the meridional flows. 
       The critical distances for stationary toroids with 
       meridional flows 
       become larger than those for stationary toroids  
       without  meridional flows. 
       In addition to this, the distances increase
       as the strengths of  meridional flows becomes larger.
       This is what we have discussed in Sec. \ref{Sec:intro}, i.e. the effects of the meridional flows 
       and the magnetic fields are oppositely working.
 \item Changing the value of the parameter $m$ in the
       certain choice of the arbitrary function $\mu$, 
       we can change the distributions of the poloidal magnetic
       fields inside the toroids. In particular, the critical 
       distances could be smaller as the value of $m$ is decreased. 
       If we adopt $M_c = 5.0 M_\odot $ and the maximum density of
       the toroid to be $\rho_\mathrm{max} = 
       1.0 \times 10^{11} \mathrm{g/cm}^3$,
       the critical distance for the $m = -1.4$ toroids becomes
       the same order as that of the ISCO of the Schwarzschild black 
       hole with mass $M_c$. For such toroids, a general relativistic 
       treatment is necessary for their correct description. 

 \item The magnetic energy for the critical
       configuration could increase as 
       the value of $m$ is decreased.
       The magnetic energy for the critical configuration with
       $m = -1.4$ is about three times larger than that for 
       the $m=0.5$ critical configuration.

 \item By obtaining stationary configurations of axisymmetric
       magnetized self-gravitating polytropic toroid with meridional
       flows under the ideal MHD approximation,
       we have shown that the effects 
       of the meridional flows would work oppositely to those of the
       poloidal magnetic fields. In other words, the oppositely working 
       effects can be easily understood if we consider that
       the dense magnetic field lines expand the gaseous 
       configurations due to the repulsive nature of
       the magnetic field lines  and that the 
       presence of the meridional flows works as lowering the
       gas pressure due to the appearance of the ram pressure
       as seen from the stationary condition equation.

\end{enumerate}

\subsection*{ACKNOWLEDGMENTS}
 
The authors would like to thank Dr. Shin. Yoshida and 
Dr. K. Taniguchi for their useful comments and discussions.
The authors would also like to thank 
the anonymous reviewer for useful comments and suggestions that helped us
to improve this paper. 
This research was supported by Grand-in-Aid for JSPS Fellows 
and for Scientific Research (24540245). 

\bibliographystyle{mn}


\appendix

\section{Integrability of basic equations and appearance of
arbitrary functions: Magnetic flux function based formulation}
\label{app:integrability}

\subsection{Integrability conditions and expression for the current density} 


Since equilibrium configurations treated in our problem are axisymmetric,
Eqs.(\ref{Eq:continuity}) and (\ref{Eq:div_B}) are automatically
satisfied by introducing the stream function $Q(R,z)$ and the
magnetic flux function $\Psi(R,z)$, defined by 
\begin{eqnarray}
   v_R \equiv  -\frac{1}{4\pi \rho R} \P{Q}{z} \ , 
      \hspace{10pt}
   v_z \equiv  \frac{1}{4 \pi \rho R} \P{Q}{R} \ ,
   \label{Eq:stream_func}
\end{eqnarray}
and
\begin{eqnarray}
   B_R \equiv -\frac{1}{R}\P{\Psi}{z} \ , 
    \hspace{10pt}
   B_z \equiv  \frac{1}{R} \P{\Psi}{R} \ .
    \label{Eq:flux}
\end{eqnarray}
If and only if the meridional flows and the poloidal magnetic fields exist simultaneously,  
the stream function is given by a function of the magnetic flux function, i.e., 
\begin{eqnarray}
 Q = Q (\Psi).
\end{eqnarray}
This relation is obtained from 
Eqs.(\ref{Eq:rot_E}) and (\ref{Eq:ideal_MHD}).   
From the $\varphi$-components of Eqs.(\ref{Eq:rot_E}) and 
(\ref{Eq:ideal_MHD}),  we obtain 
\begin{eqnarray}
    {v_{\varphi} 
      - \displaystyle{Q' \over 4 \pi \rho} B_{\varphi} \over R} 
    = \Omega(\Psi) \ ,
\end{eqnarray}
where $^{'}$ means the derivative with respect to $\Psi$ and 
$\Omega(\Psi)$ is an arbitrary function of $\Psi$.
From integrability conditions for Eq. (\ref{Eq:motion_tor}), 
we obtain two relations: 
\begin{eqnarray}
   R B_\varphi - Q'(\Psi) R v_\varphi = \kappa(\Psi) \ , 
\end{eqnarray}
{\footnotesize
\begin{eqnarray}
   -\frac{B_\varphi \kappa'(\Psi) }{4 \pi \rho R} 
      - R v_\varphi \Omega'(\Psi)
      - \frac{v_\varphi Q''(\Psi) B_\varphi}{4 \pi \rho } 
      + \frac{j_\varphi/c}{\rho R} 
      - \frac{\omega_\varphi Q'(\Psi)}{4 \pi \rho R} 
    = \mu(\Psi) \ ,
\label{Eq:mu_def}
\end{eqnarray}
}
where $\kappa(\Psi)$ and $\mu(\Psi)$ are arbitrary functions of $\Psi$.
Using  Eqs. ({\ref{Eq:rot_H}) and (\ref{Eq:mu_def}), we may  
describe the current density $\Vec{j}$ 
as follows:
\begin{eqnarray}
 \frac{\Vec{j}}{c} &= \left[\kappa'(\Psi) 
       + R v_\varphi Q''(\Psi)  \right] \frac{\Vec{B}}{4\pi}
       + Q' (\Psi)  \frac{\Vec{\omega}}{4\pi} \nonumber \\
       &+ \rho R \left[\mu(\Psi) 
       + R v_\varphi \Omega'(\Psi) \right]\Vec{e}_\varphi \ .
   \label{Eq:current}
\end{eqnarray}

It should be noted that in this current density formula, 
the four arbitrary functions appear and that each function is
related to different physical quantities 
corresponding to each term in the right-hand side of Eq. (\ref{Eq:current}).
\cite{Lovelace_et_al_1986} obtained the current density 
including another arbitrary function which is related to 
the entropy. This arbitrary function, related to the entropy,  
does not appear in our problem because uniform entropy distributions is implicitly assumed.

\subsection{Choice of arbitrary functions in this study}
\label{app:arbitrary_functions}

We choose the functions $\kappa$,  $Q'$, $\Omega$
and $\mu$ as:
\begin{eqnarray}
  \kappa(\Psi) =
   \left\{
     \begin{array}{lr}
        0 \ , & \mathrm{for} \hspace{10pt} \Psi \leq \Psi_{\max} \ , \\
        \dfrac{\kappa_0}{k+1}(\Psi - \Psi_{\max})^{k+1} \ , & \mathrm{for} \hspace{10pt} \Psi \geq \Psi_{\max} \ ,
     \end{array}
   \right.
   \label{Eq:kappa}
\end{eqnarray}
\begin{eqnarray}
  Q'(\Psi) =
   \left\{
     \begin{array}{lr}
        0 \ , & \mathrm{for} \hspace{10pt} \Psi \leq \epsilon_Q \Psi_{\max} \ , \\
        \dfrac{Q_0}{k + 1}(\Psi - \epsilon_Q \Psi_{\max})^{k+1} \ , & \mathrm{for} \hspace{10pt} \Psi \geq \epsilon_Q \Psi_{\max} \ ,
     \end{array}
   \right.
   \label{Eq:Q}
\end{eqnarray}
\begin{eqnarray}
   \Omega(\Psi) & = & \Omega_0(\Psi^2 + d^2)^{\alpha} \ , \\
   \mu(\Psi) & = & \mu_0 (\Psi + \epsilon )^m \ , 
     \label{Eq:mu}
\end{eqnarray}
where, $\kappa_0$, $k$, $Q_0$, $\epsilon_Q$, $\Omega_0$,
$d$, $\mu_0$, $\epsilon$, $\alpha$, and $m$ are constant parameters 
and $\Psi_{\max}$ denotes the 
maximum value of $\Psi$ in the vacuum region. 
This choice of $\kappa$ is the same as that employed in \cite{Tomimura_Eriguchi_2005}, 
\cite{Yoshida_Eriguchi_2006},
\cite{Yoshida_Yoshida_Eriguchi_2006}, and
\cite{Otani_Takahashi_Eriguchi_2009}.
We introduce the parameter $\epsilon_Q$ in Eq. (\ref{Eq:Q}) 
and set $\epsilon_Q = 1.1$ in order to restrict the region where 
meridional flows exist well inside the surface of the toroid. 
Choosing these functional forms, we can 
avoid singular behavior of the solutions which could appear on the surface of the toroid. 

\section{Stream function based formulation: Vorticity formula}
\label{app:integrability2}

So far, we consider the situations in which the poloidal magnetic fields exist 
everywhere except in vacuum  region. For such situations, as already shown, the flux function 
$\Psi$ can be a principal variable by which all the magnetohydrodynamical  
quantities are determined. 
If we assume that the poloidal velocity 
fields exist  everywhere inside the fluid region, 
the same problem as that treated in this study may be formulated
by considering the stream function $Q$ as a principal variable, which is named ``the stream function based formulation''. 
For this formulation, the magnetic flux function is given by a function of the stream function, 
\begin{eqnarray}
    \Psi = \Psi(Q) \ .
\end{eqnarray}
The other arbitrary functions of the magnetic flux function defined 
in Appendix \ref{app:integrability} 
are regarded as functions of $Q$.
Then, the vorticity vector may be written as:
\begin{eqnarray}
  \Vec{\omega} &=& 4 \pi \rho \left\{{d \ell(Q) \over dQ} + 
       R {d^2 \Psi(Q) \over d Q^2}  
      B_{\varphi} \right\} \Vec{v} \nonumber \\
     &+& 4 \pi {d \Psi(Q) \over d Q} \frac{\Vec{j}}{c} 
     + \rho R \left\{ - \nu(Q) + R 
        \displaystyle{d \sigma(Q) \over dQ} B_{\varphi} \right\}
        \Vec{e}_{\varphi} \ ,
\end{eqnarray}
where $\ell(Q)$, $\nu(Q)$ and $\sigma(Q)$ are
another set of arbitrary functions of $Q$ for the stream function based formulation.
The arbitrary functions $\ell(Q)$ and $\sigma(Q)$ are related to 
the physical quantities $v_{\varphi}$ and $B_{\varphi}$ through 
\begin{eqnarray}
  \ell(Q) = R v_{\varphi} - R  {d \Psi(Q) \over dQ} 
   B_{\varphi} \ ,
\end{eqnarray}
\begin{eqnarray}
  \sigma(Q) = {B_{\varphi} \over \rho R}  
        - { 4 \pi \over R} v_{\varphi}
              {d \Psi(Q) \over d Q} \ .
\end{eqnarray}
%

\section{Physical quantities}
\label{app:physical_quantities}

\subsection{Global physical quantities}

Some useful global quantities are defined as follows:
The gravitational energy of the magnetized toroid and the central object 
is defined by 
\begin{eqnarray}
   W \equiv \frac{1}{2}\int \left(\phi_g - 
\frac{G M_c}{r} \right) \rho \, d^3 \Vec{r} \ , 
\end{eqnarray}
where the self energy of the central object is discarded. 
The kinetic energy of the fluid is defined by  
\begin{eqnarray}
   T \equiv \frac{1}{2} \int \rho |\Vec{v}|^2 \, d^3 \Vec{r} \ .
\end{eqnarray}
The volume integral of the pressure is defined by 
\begin{eqnarray}
   \Pi \equiv \int p \, d^3 \Vec{r} \ .
\end{eqnarray}
Through $\Pi$, the internal energy is defined as
\begin{eqnarray}
 U \equiv N \Pi \, .
\end{eqnarray}
The magnetic energy is defined by  %
\begin{eqnarray}
\mathfrak{M}  \equiv \int r \cdot \left(\frac{\Vec{j}}{c}
 \times \Vec{B} \right) \, d^3 \Vec{r} \,.
 \label{Eq:M_energy}
\end{eqnarray}
The mass of the toroid is defined by 
\begin{eqnarray}
   M \equiv \int \rho d^3 \Vec{r} \ .
\end{eqnarray}
In order to evaluate the effect of the meridional flows,
we also define  the kinetic energy of the
meridional flows as follows:
\begin{eqnarray}
 T_p \equiv \frac{1}{2} \int \rho v_p^2 \, d^3 \Vec{r}.
\end{eqnarray}
%

\subsection{Dimensionless physical quantities}

\label{app:dimensionless}

For the numerical computations, the physical quantities are 
transformed into dimensionless ones. 
The dimensionless quantities employed in this study are defined as follows:

\noindent
For local quantities, 
\begin{eqnarray}
   \hat{r} \equiv \frac{r}{r_e} = \frac{r}{
   \sqrt{\frac{1}{\beta}\frac{p_{\mathrm{max}}}{4\pi G
   \rho_{\mathrm{c}}^2}}} \, \hspace{10pt} , 
\label{Eq:r_dim}
\end{eqnarray}
\begin{eqnarray}
   \hat{\rho} \equiv \frac{\rho}{\rho_\mathrm{c}} \ ,
\end{eqnarray}
\begin{eqnarray}
   \hat{\phi}_g \equiv \frac{\phi_g}{4 \pi G r_e^2 \rho_{\mathrm{c}}} \ ,
\end{eqnarray}
\begin{eqnarray}
   \hat{\Omega} \equiv \frac{\Omega}{\sqrt{{4 \pi G 
       \rho_{\mathrm{c}}}}} \ ,
\end{eqnarray}
\begin{eqnarray}
   \hat{\Vec{v}} \equiv \frac{\Vec{v}}{\sqrt{{4 \pi G r_e^2 
       \rho_{\mathrm{c}}}}} \ ,
\end{eqnarray}
\begin{eqnarray}
 \hat{\kappa} \equiv \frac{\kappa}{\sqrt{4\pi G} r_e^2 \rho_{\rm{c}}} \ ,
\end{eqnarray}
\begin{eqnarray}
 \hat{\mu} \equiv \frac{\mu}{\sqrt{4 \pi G}/r_e} \ ,
\end{eqnarray}
\begin{eqnarray}
   \hat{\Vec{B}} \equiv \frac{\Vec{B}}{\sqrt{4 \pi G }r_e 
     \rho_{\rm{c}} } \ ,
\end{eqnarray}
\begin{eqnarray}
   \hat{A}_\varphi \equiv \frac{A_\varphi}{\sqrt{4 \pi G }r_e^2 
      \rho_{\rm{c}} } \ ,
\end{eqnarray}
\begin{eqnarray}
 \hat{\Psi} \equiv \frac{\Psi}{\sqrt{4 \pi G }r_e^3 \rho_{\rm{c}} } \ ,
\end{eqnarray}
\begin{eqnarray}
   \hat{K} \equiv \frac{K}{4 \pi G r_e^6 \rho_{\rm{c}}^2 } \ ,
\end{eqnarray}
\begin{eqnarray}
   \hat{j}_\varphi \equiv \frac{j_\varphi}{\sqrt{4\pi G}
      \rho_{\rm{c}} c} \ .
\end{eqnarray}
\begin{eqnarray}
   \hat{C} \equiv \frac{C}{4 \pi G r_e^2 \rho_{\rm{c}}} \ ,, 
\end{eqnarray}
where $r_e$ and $p_{\max}$ denote the equatorial radius of 
the outer edge of the toroid and 
the maximum pressure $p_{\max}$,  respectively.

\noindent 
As for the global quantities, 
\begin{eqnarray}
   \hat{M} = \frac{M}{r_e^3 \rho_{\rm{c}}} \ ,
\end{eqnarray}
\begin{eqnarray}
   \hat{W} = \frac{W}{4\pi G r_e^5 \rho_{\rm{c}}^2} \ ,
\end{eqnarray}
\begin{eqnarray}
   \hat{T} = \frac{T}{4\pi G r_e^5 \rho_{\rm{c}}^2} \ ,
\end{eqnarray}
\begin{eqnarray}
   \hat{\Pi} = \frac{\Pi}{4\pi G r_e^5 \rho_{\rm{c}}^2} \ ,
\end{eqnarray}
\begin{eqnarray}
 \hat{{\mathfrak M}} = \frac{{\mathfrak M}}{4\pi G r_e^5 \rho_{\rm{c}}^2} \ .
\end{eqnarray}
Here, $\beta$ appearing in Eq. (\ref{Eq:r_dim}) is introduced so as to make the distance from the symmetry axis to the outer edge of
the toroid to be unity (or $\hat{r}=1$) during numerical iterations (see, e.g., \cite{Otani_Takahashi_Eriguchi_2009}).
Thus, values of $\beta$ are obtained after converged solutions are obtained.
Increasing $\beta$ results in decrease in 
$r_e$ and decreasing $\beta$ results in
increase in $r_e$.

Using the dimensionless quantities defined above, we obtain the dimensionless
form of the right-hand side of Eq. (\ref{eq:first_integral}),
\begin{eqnarray}
 \frac{1}{4 \pi G r_e^2 \rho_{c}}\int \frac{d p}{\rho} &=& -\hat{\phi}_g 
 + \frac{\hat{M}_c}{4 \pi \hat{r}}  - \frac{1}{2}|\hat{\Vec{v}}|^2 
+ \int_{\hat{\Psi}} \hat{\mu}(\hat{\Psi}) \, d \hat{\Psi} \nonumber \\ 
&+& \hat{r} \sin \theta \hat{v}_\varphi \hat{\Omega}(\hat{\Psi}) 
+ \hat{C}.
\label{Eq:6_first_integral}
\end{eqnarray}
We integrate the left side 
of this equation by using polytrope relation
as follows:
\begin{eqnarray}
 \frac{1}{4\pi G r_e^2 \rho_{c}}\int \frac{d p}{\rho} 
  = \beta (N + 1) \hat{\rho}^{1/N}.
\end{eqnarray}
Then, we obtain the dimensionless form of Bernoulli's equation as follows:
\begin{eqnarray}
\beta (N + 1) \hat{\rho}^{1/N} &=& -\hat{\phi}_g 
 + \frac{\hat{M}_c}{4 \pi \hat{r}}  - \frac{1}{2}|\hat{\Vec{v}}|^2 
+ \int_{\hat{\Psi}} \hat{\mu}(\hat{\Psi}) \, d \hat{\Psi} \nonumber \\
&+& \hat{r} \sin \theta \hat{v}_\varphi \hat{\Omega}(\hat{\Psi}) 
+ \hat{C}.
\label{App:first_integral}
\end{eqnarray}
We have used this dimensionless forms in actual numerical computations.

\section{Numerical method}
\label{app:scheme_method}

\begin{figure}
 \begin{center}
 \includegraphics[scale=1.2]{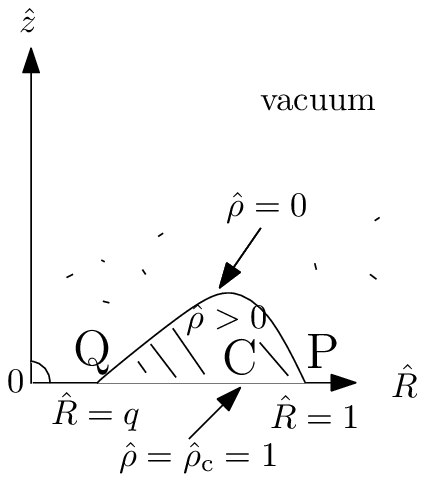}
  \caption{Three points used
in the HSCF scheme.}
  \label{Fig:HSCF}
 \end{center}
\end{figure}

In our numerical studies, the generalized iteration scheme
known as  Hachisu's Self-Consistent-Field (HSCF) method 
(\citealt{Hachisu_1986a}) is adopted in order to solve the 
system of non-linear partial differential equations for 
equilibrium configurations of magnetized toroids. 
In  this generalized HSCF method, the density, the gravitational potential
and the vector potential are discretized on grid points  $(r_i, \theta_j)$ and non-linear algebraic 
equations for these discretized physical quantities are iteratively solved (\citealt{Tomimura_Eriguchi_2005}).

We start our computation by assuming initial guesses
for the discretized mass density and flux function.
Using a trial density and a flux function distribution, the
dimensionless gravitational potential $\hat{\phi}_g$ and the
vector potential $\hat{A}_{\varphi}$ 
are respectively computed through the integrations, which are equivalent to 
Eqs.  (\ref{Eq:2_Poisson_int}) and (\ref{Eq:A_phi}), 
\begin{eqnarray}
   \hat{\phi}_g(\hat{r},\theta)
    &=& - \sum_{n=0}^{n_{\max}} \mathrm{P}_{2n}(\cos\theta) 
\int_0^\infty d\hat{r}' \hat{r}'^2
                  f_{2n}(\hat{r},\hat{r}')
\nonumber \\
&\times&    \int_0^{\pi/2} d\theta'\sin \theta'
                \mathrm{P}_{2n}(\cos\theta') \hat{\rho}(\hat{r}',\theta' ),
	\label{Eq:non-d-gphi}
\end{eqnarray}
\begin{eqnarray}
   \hat{A}_\varphi (\hat{r},\theta)
    &=&  4\pi \sum_{n=1}^{n_{\max}}
                 \frac{ \mathrm{P}_{2n-1}^1(\cos\theta)}{2n(2n-1)}
                 \int_0^\infty d \hat{r}' \hat{r}'^2
                   f_{2n-1}(\hat{r}, \hat{r}')
\nonumber \\
&\times&    \int_0^{\pi/2} d\theta' \sin\theta'
      \label{Eq:non-d-Aphi}
               \mathrm{P}_{2n-1}^1(\cos\theta') \hat{j}_\varphi(\hat{r}',\theta'),
\end{eqnarray}
\begin{equation}
   f_{2n}(\hat{r},\hat{r}')
    = \left\{
                 \begin{array}{lr}
                    \hat{r}'^{2n}/\hat{r}^{2n+1} , & (\hat{r} \geq \hat{r}'), \\
                    \hat{r}^{2n}/\hat{r}'^{2n+1} , & (\hat{r} \leq \hat{r}').
                 \end{array}
               \right.
\end{equation}
Here P$_n (\cos \theta)$ are the Legendre polynomials of degree $n$  and P$^1_n (\cos \theta)$ 
are the associated Legendre functions of order one. In the actual computations, we take
$n_{\max} = 40$ typically. We use Eq. (\ref{Eq:current}) to obtain the 
distribution of $\hat{j}_\varphi$. After these integrals, we calculate 
the distribution of the flux function using the relation of Eq. (\ref{Eq:Psi_Aphi}).
Although there are many constant parameters in our formulation, three constants,  
the length scale factor $\beta$, the integration constant $\hat{C}$, and the strength of 
the toroidal current $\hat{\mu}_0$ 
are especial in the sense that they cannot be specified before converged solutions 
are obtained. These three constants are determined by imposing the following three
conditions given at three special grid points and therefore change their values 
during the iteration procedures.  
Fig. \ref{Fig:HSCF} shows a schematic image of our scheme.
At the inner surface (point {\bf Q}), 
i.e. at $ \hat{r} = q$ and $\theta = \pi/2$, and 
at the outer surface (point {\bf P}), i.e. at
$\hat{r} = 1$ and $\theta = \pi/2$, the density must vanish.
At the point where the density takes its maximum value (point C), i.e. at
a point of $\hat{r} = \hat{r}_\mathrm{max}$ and 
$\theta = \theta_\mathrm{max}$, the dimensionless 
density must be unity. From Bernoulli's equation (\ref{App:first_integral})
and three conditions are explicitly given by
\begin{eqnarray}
   0 = - {\hat{\phi}_g}|_P + \frac{\hat{M}_c}{4 \pi \hat{r}}\Big|_P   - 
\frac{1}{2}({\hat{v}_r}^2 + {\hat{v}_\theta}^2 + {\hat{v}_\varphi}^2)|_P
\nonumber \\ 
             + \frac{\hat{\mu}_0}{m + 1} (\hat{\Psi}|_P + \epsilon)^{m+1} 
+\hat{r} \sin \theta \hat{v}_\varphi \hat{\Omega}_0 (\hat{\Psi}^2 + d^2)^\alpha|_P + \hat{C} \ ,
\end{eqnarray}
\begin{eqnarray}
   0 = 
- {\hat{\phi}_g}|_Q + \frac{\hat{M}_c}{4\pi\hat{r}}\Big|_Q  - 
\frac{1}{2}({\hat{v}_r}^2 + {\hat{v}_\theta}^2 + {\hat{v}_\varphi}^2)|_Q
\nonumber \\ 
             + \frac{\hat{\mu}_0}{m + 1} (\hat{\Psi}|_Q + \epsilon)^{m+1} 
+\hat{r} \sin \theta \hat{v}_\varphi \hat{\Omega}_0 (\hat{\Psi}^2 + d^2)^\alpha|_Q + \hat{C} \ ,
\end{eqnarray}
\begin{eqnarray}
   \beta (1+N) =
- {\hat{\phi}_g}|_C + \frac{\hat{M}_c}{4\pi\hat{r}}\Big|_C  - 
\frac{1}{2}({\hat{v}_r}^2 + {\hat{v}_\theta}^2 + {\hat{v}_\varphi}^2)|_C
\nonumber \\ 
+ \frac{\hat{\mu}_0}{m + 1} (\hat{\Psi}|_C + \epsilon)^{m+1} 
+\hat{r} \sin \theta \hat{v}_\varphi \hat{\Omega}_0 (\hat{\Psi}^2 + d^2)^\alpha|_C + \hat{C} \ ,
\end{eqnarray}
Solving these equations, we obtain three constants $\beta$, $\hat{C}$ and $\hat{\mu}_0$.
Using the three constants, gravitational potential and flux function obtained newly,
 we solve Bernoulli's equation in terms of the matter density as follows:
{\small
\begin{eqnarray}
  \hat{\rho} =  \Big(
\dfrac{- {\hat{\phi}_g} + \frac{\hat{M}_c}{4\pi\hat{r}}  - 
\frac{1}{2}({\hat{v}_r}^2 + {\hat{v}_\theta}^2 + {\hat{v}_\varphi}^2)
             + \frac{\hat{\mu}_0}{m + 1} (\hat{\Psi} + \epsilon)^{m+1} 
+\hat{r} \sin \theta \hat{v}_\varphi \hat{\Omega}_0 (\hat{\Psi}^2 + d^2)^\alpha + \hat{C}}{ \beta (1 + N)}
\Big)^{N} \ .
\end{eqnarray}
}

The newly obtained density and other quantities are used as
a new guess for the next potentials in the iteration cycle. We carry out
this iteration procedure until the relative changes of all physical
quantities between two iteration cycles becomes less than some
prescribed small number, e.g., $10^{-4}$.

\section{Numerical accuracy check }
\label{app:Numerical}

\begin{figure*}
 \includegraphics[scale=0.6]{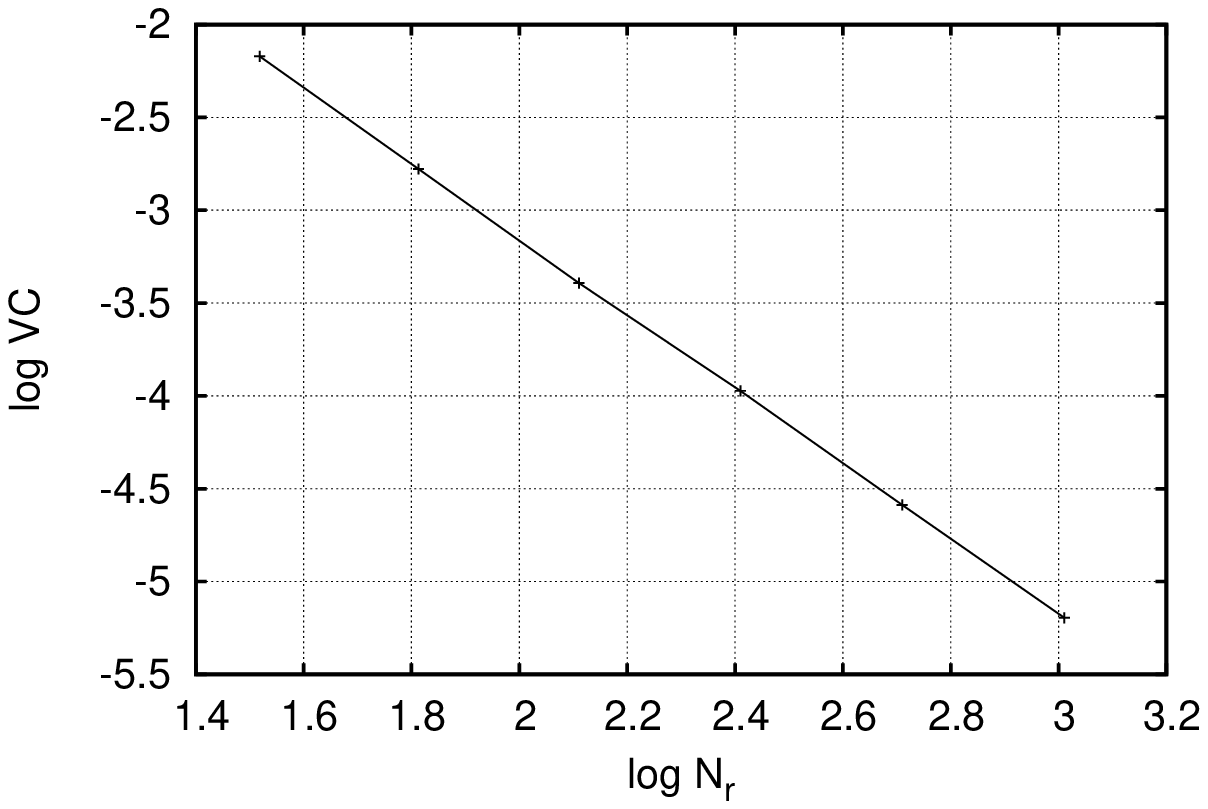}
\caption{Values of VC vs. the number of 
the grid points in the $r$ direction.
}
\label{Fig:VC}
\end{figure*}

In order to check the convergence and the accuracy
of the obtained stationary solutions, 
we use a virial relation, 
\begin{eqnarray}
 2T + W + 3 \Pi + \mathfrak{M} = 0.
\label{Eq:VC}
\end{eqnarray}
This relation is satisfied for the exact stationary solutions.
Numerical solutions, however, have  
numerical errors. Thus, in general, 
the right-hand side of Eq. (\ref{Eq:VC}) does not vanish  
although its value should be sufficiently small.
In order to quantify numerical errors
contained in the numerical solutions obtained, 
we evaluate a virial relation, defined by 
 \begin{eqnarray}
  {\rm VC} =  \left|\frac{2\hat{T} + \hat{W} + 3 \hat{\Pi} + \mathfrak{\hat{M}}}{\hat{W}}\right|.
 \end{eqnarray}
VC is  an estimate of the numerical errors and  has been used in many papers 
(e.g., \citealt{Hachisu_1986a}; \citealt{Tomimura_Eriguchi_2005}; 
\citealt{Lander_Jones_2009}; \citealt{Otani_Takahashi_Eriguchi_2009}; 
\citealt{Fujisawa_Yoshida_Eriguchi_2012}).
If a value of VC is less than $10^{-4}$ - $10^{-5}$, the solution is considered to be 
sufficiently accurate. We have checked the 
convergence of VC by changing the gird number 
in the $\hat{r}$ direction, $N_r$, 
to test our numerical code in which
we have  used the spherical coordinates. 
As for the grid number in the $\theta$ 
direction, $N_\theta$, 
we fix it as $N_\theta = 129$ during the test
computations.
Since we assume equatorial symmetry and 
the gravitational potential of the central object has a singularity 
at $\hat{r}=0$, 
we consider the region defined by 
$\hat{r} \in [1.0 \times10^{-2}:4.0]$ and 
$\theta\in[0:\pi/2]$ as our computational space.
In order to resolve structures of the toroid 
properly, 
we use non-uniformly distributed grid points used in \citealt{Fujisawa_Yoshida_Eriguchi_2012}). 
We compute equilibrium configurations with 
$q=0.6$, $M_t/M_c = 2.0 \times 10^{-2}$, $m=0$, 
$\hat{\kappa}_0 =4.5$, $\hat{Q}_0=0.0$,
$\hat{\Omega}_0 = 0.0$ (no rotation) for 
several different values of $N_r$.
Fig. \ref{Fig:VC} shows values of log VC against values of log $N_r$.
In this figure, we observe that as the grid number in the $r$-direction is 
increased, the value of VC becomes
smaller as 
\begin{eqnarray}
 {\rm VC} \propto N_r^{-2}\, .
\end{eqnarray}
This behavior of VC is quite reasonable and proper because we use the numerical scheme with
the second-order accuracy formula for numerical derivatives and Simpson's integration formula. 
If $N_r = 513$ ($\log N_r \sim 2.710$),
the value of VC is $\sim 10^{-5}$ ($\log$ VC $
= -4.58$). In the present study, we choose the grid number as $N_r = 513$ and 
$N_\theta = 513$ because this grid number is sufficiently large to obtain numerical 
solutions  with acceptable accuracy as shown before. 

\bsp

\end{document}